# Natural polyphenols inhibit the dimerization of the SARS-CoV-2 main protease: the case of fortunellin and its structural analogs

Athanasios A. Panagiotopoulos[*], Danai-Maria Kotzampasi[*], George Sourvinos[†,#], Marilena Kampa[*,#], Stergios Pirintsos[‡,&,#], Elias Castanas[*,#,‖], Vangelis Daskalakis[¶,‖]

* Laboratory of Experimental Endocrinology, University of Crete, School of Medicine, Heraklion, Greece

† Laboratory of Virology, University of Crete, School of Medicine, Heraklion, Greece

‡ Department of Biology, University of Crete, 71409 Heraklion, Greece

& Botanical Garden, University of Crete, Rethymnon, Greece

¶ Department of Chemical Engineering, Cyprus University of Technology, Limassol, Cyprus

# Nature Crete Pharmaceuticals, Heraklion, Greece

‖ Corresponding Authors

**Address all correspondence to:**

Dr Elias Castanas, University of Crete, School of Medicine, Voutes University Campus, Heraklion, 71013, Greece (castanas@uoc.gr)

Dr Vangelis Daskalakis, Department of Chemical Engineering, Cyprus University of Technology, Limassol, 3041, Cyprus (evangelos.daskalakis@cut.ac.cy)

**Abstract**

3CL-Pro (or M-Pro) is the SARS-CoV-2 main protease, acting as a homodimer, is responsible for the cleavage of the large polyprotein 1ab transcript in proteins acting on viral growth and replication. 3CL-Pro has been one of the most studied SARS-CoV-2 proteins and the subject of therapeutic interventions, targeting its catalytic domain. A number of drug candidates have been reported, including some natural products. Here, we investigated *in silico*, through binding and molecular dynamics simulations, the natural product space for the identification of candidates of 3CL-Pro dimerization inhibitors. We report that fortunellin (acacetin 7-O-neohesperidoside), a natural flavonoid O-glycoside, is a potent inhibitor of 3CL-Pro dimerization. A search of the ZINC natural products database identified another 16 related molecules, including apilin and rhoifolin, with interesting pharmacological properties. We propose that fortunellin and its structural analogs might be the basis of novel pharmaceuticals and dietary supplements against SARS-CoV-2 induced COVID-19 disease. Our findings are supported by the experimental literature.

## Introduction

Severe acute respiratory syndrome coronavirus 2 (SARS-CoV-2) was first identified in December 2019, and is the causative agent of coronavirus disease 2019 (COVID-19). It became a global pandemic, threatening the lives of millions of people belonging to sensitive health groups. Intense scientific effort worldwide resulted in the identification of SARS-CoV-2 genomic structure, viral protein sequence and structure and disease characteristics.[1,2] 3CL-Pro (or Mpro), the main protease of SARS-CoV-2, plays a key role in polyprotein processing. This enzyme plays a vital role in cleaving the large polyprotein 1ab (replicase 1ab, ~790 kDa) translated by the virus RNA, at 11 different sites, liberating proteins indispensable for viral replication and proliferation, with a unique specificity, not found in any human protease.[3] Therefore, it represents a preferential target for the development of a series of not toxic inhibitors. The individual monomers of SARS-CoV-2 3CL-Pro are inactive; thus, inhibitors of its dimerization are needed. The chemical space of natural products, and especially that of polyphenols, has provided valuable candidate lead molecules in a number of diseases, including COVID-19.[4] Here, we have interrogated this resource, in view of identifying potential candidates inhibiting the dimerization of SARS-CoV-2 3CL-Pro. Herein, Fortunellin (acacetin 7-O-neohesperidoside), a natural flavonoid O-glycoside, isolated from the fruits of *Citrus japonica* var. *margarita* (kumquat),[5] was found as a lead compound, which, together with a series of 16 structurally related analogs, including apilin and rhoifolin, could be used as the basis for the design of novel antiviral compounds.

## Material and Methods

The Swiss Model Biospace contains 110 crystal structures of 3CL-Pro (July 2020), bound or not to ligands, and deposited to the Protein Data Bank (https://www.rcsb.org/) with high structural similarity (**Table S1** in the Supplemental Material). Here, we base our docking study on the crystal with reference 6YB7, in the absence of a ligand. For the study, ligands and their analogs were retrieved from the ZINC database (http://zinc.docking.org/).[6] Fully flexible ligand binding on 3CL-Pro by the GalaxyWeb server (http://galaxy.seoklab.org/), was performed as described previously.[7] Molecular Dynamics (MD) studies were based on two unligated crystals (6YB7, 6LU7) [8] to increase phase space sampling. Classical MD simulations of the 3CL-Pro monomer (60μs) described by the Amber03 force field were performed in GROMACS 2020. Retrieved trajectories were further analyzed by Markov state modeling (MSM).[9] Time-structure independent components analysis (tICA) was used to reduce the dimensionality of our data, in PyEMMA. MSM gave the conformational phase space of MPro, in two Collective Variables (CV-1/ CV-2) determined by the torsional angles of residues 3, 4, 5, 6, 84, 135, 141, 164, 167, 171, 175, 178, 179, 180, 190, 195, 217, 284, 285, 286, 290, 291, 300, and 301. Please, refer to the Supplemental Material for a detailed description of the methods used in this study and the detailed conformational space functions out of the MSM analysis.

**Results and Discussion**

The Classical MD of the homodimer of 3CL-Pro (10μs), followed by 20μs enhanced sampling (parallel tempering metadynamics at the well-tempered ensemble, PTmetaD-WTE method)[10], on the above CV phase space, revealed three conformations (**Figure 1A**), at the minima of the free energy surface (FES). The first conformation (C1) corresponds to the crystal structure of the dimer (monomer distance 1.72-1.78 Å), while the other two (C2-C3) correspond to looser dimer structures with higher (1.86-1.93 Å) distances among monomers, resulting in weakened monomer-monomer interactions. The monomer distances have been calculated based on the minimum distances between residues on the monomer-monomer interface (4, 10, 11, 14, 28, 139, 140, 147, 290, 298).[11]

We utilized the FTMap server [12] to identify binding hot spots, determine drugability and provide information about fragment-based drug discovery on 3CL-Pro. The results of FTMap permit us to design a minimal structure of a potential 3CL-Pro dimerization inhibitor. Interrogating the ZINC database of natural products with this minimal structure, we have identified fortunellin (ZINC4349204,

**Figure 1B**) as a potential natural inhibitor of 3CL-Pro dimerization. Binding of fortunellin on the 3CL-Pro monomer (**Figure 1C**) revealed a high affinity binding (ΔG -13.9 kcal/mol) using the fully flexible GalaxyWeb server, and strong interactions with 3CL-Pro amino acids ($Leu_{32}$, $Asp_{33}$, $Asp_{34}$, $Val_{35}$, $Tyr_{37}$, $Gln_{83}$, $Lys_{88}$, $Tyr_{101}$, $Lys_{102}$, $Phe_{103}$, $Val_{104}$, $Arg_{105}$, $Asp_{108}$, $Phe_{159}$, $Cys_{160}$, $Asp_{176}$, $Leu_{177}$ and $Glu_{178}$). Despite the fact that fortunellin does not bind at the dimerization interface, as defined by the residues 4, 10, 11, 14, 28, 139, 140, 147, 290, and 298, the binding might allosterically inhibit, or weaken the formation of the active dimer, as indicated by the large-scale classical and enhanced sampling MD simulations (*see below*). To increase the statistical significance of the docking studies, we have also used the different poses of the monomeric 3CL-Pro MD solutions for 60 μs, at 1 ns interval (poses were retrieved every 3 μs) as scaffolds for fortunellin binding. As shown, the affinity of fortunellin is fluctuating around -12.3 kcal/mol (SD ±1.005 kcal/mol), and, after initial fluctuations, preceding significant changes of the dimerization domain, as expressed by changes of the local RMSD value, as compared to the crystal structure of 3CL-Pro, it stabilizes after 30 μs (see Supplemental Material **Table S2**).

In **Figures 1D-E**, we present the MSM analysis of the equilibrium dynamics of MPro in the presence and absence of fortunellin, respectively on the MSM refined CV phase space (CV-1/ CV-2). The C1-C3 minima are assigned in these graphs based on structural feature comparison with **Figure 1A**. Based on MSM, we calculated that the transitions between the different states (C1-C3) occur at the average time scales of 37.6ns (slow processes) and  3.5ns (fast processes) in the absence of fortunellin (**Figure 1F**). However, the average transition time scale in the presence of fortunellin drops to 1.42ns with only fast transitions (**Figure 1G**). This significantly smaller time-interval results in greater energy changes faster in the presence of fortunellin, blocking the trapping at certain states, and inhibiting the formation of the dimer by sampling non-favorable monomer conformations, i.e. with greater distance among monomers, that favor its dissociation. We note that the C1 minimum that is associated with the crystal structure (active for dimerization) in **Figure 1D**, it is absent in **Figure 1E**, indicating that fortunellin disfavors this state. Instead, C1 is replaced by an alternative protein state, indicated by the purple area in **Figure 1E**.

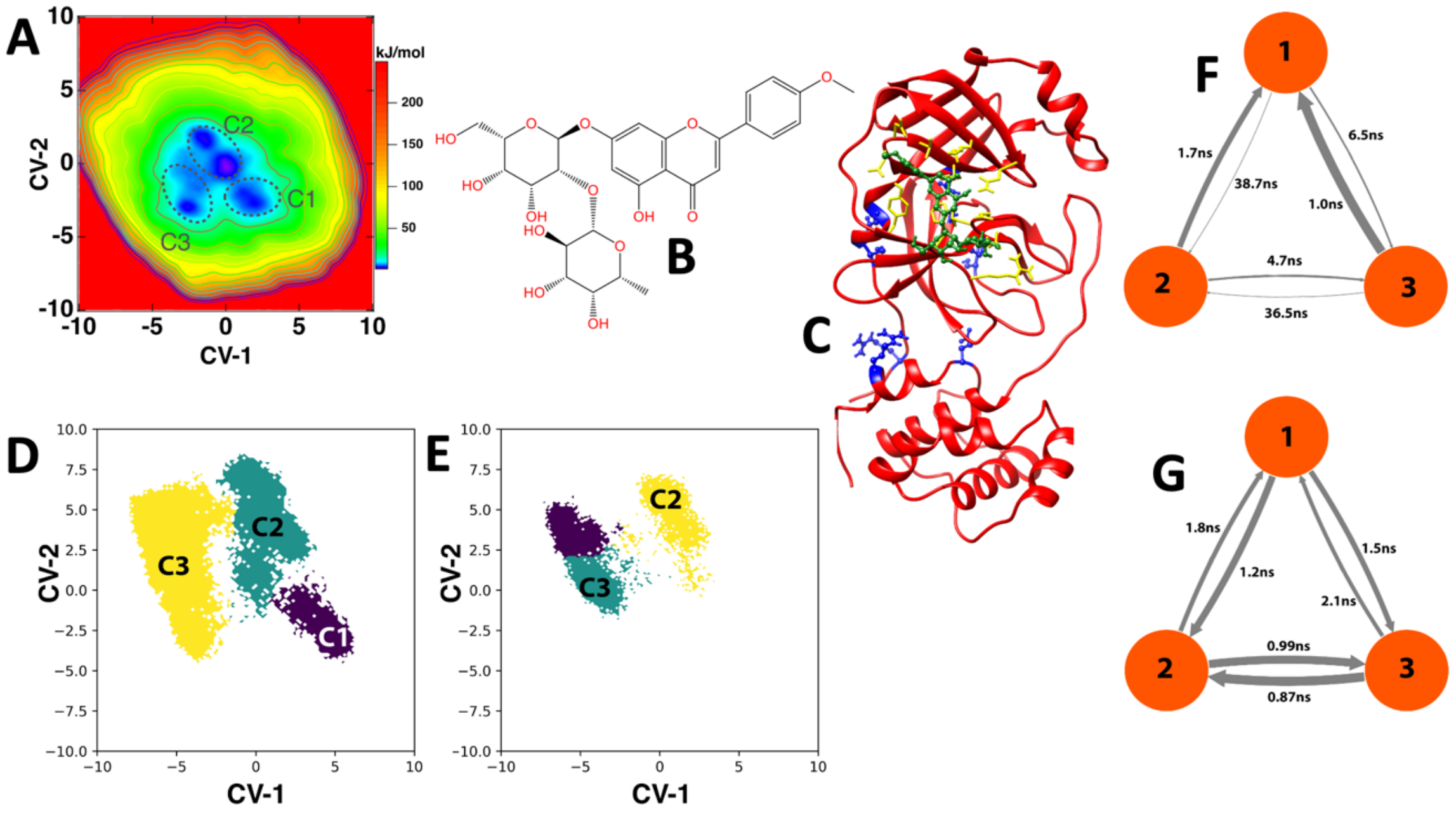

**Figure 1** | **A.** The Free Energy Surface (FES) out of the PTmetaD-WTE enhanced sampling runs. Three minima of the 3CL-Pro dimer dynamics are identified at the blue regions (C1, C2, and C3) on the CV-1/ CV-2 phase space. **B.** The molecular structure of fortunellin. **C.** The interaction of fortunellin with the 3CL-Pro monomer. Fortunellin is shown in green, the interacting amino acids are shown in yellow and the dimerization interacting aminoacids are shown in blue. **D.** The identified states of 3CL-Pro in the absence of fortunellin within CV-1, CV-2 phase space. **E.** The identified states of 3CL-Pro in the presence of fortunellin within the refined CV-1, CV-2 phase space. The C1 to C3 minima are indicated on the D-E graphs based on the comparison with Figure 1A. **F-G.** The transition times between the C1 to C3 minima are calculated in the absence (**F**) and in the presence (**G**) of fortunellin.

The 3CL-Pro C44-P52 loop has been proposed to host mutations, however only at the T45, S46, E47, L50 positions.[13] The latter are not listed as important residues in the 3CL-Pro/ fortunellin dynamics analyzed by MSM (see Supplemental Material). This gives us confidence that fortunellin can target 3CL-Pro, even when 3CL-Pro is mutated. It appears therefore that the natural polyphenol fortunellin is a good drug (or dietary supplement) candidate for combatting COVID-19 disease. The bulk of the scientific effort targeting 3CL-Pro focuses on the discovery or repurposing of inhibitors of its enzymatic activity. [14]. Here, we have instead directed our efforts toward the identification of (natural) compounds, which could inhibit the dimerization of the enzyme. Previous studies have implicated fortunellin as an activator of anti-oxidant enzymes (HO-1, SOD and CAT), through a direct action on Nrf2 and AMPK pathways, considered as important to protect against oxidative stress.[15] In addition, fortunellin was implicated as a cardioprotective factor in diabetic animals.[16] Here, we extend the actions of this agent, by reporting a direct effect of fortunellin, impairing the dimerization of 3CL-Pro SARS-CoV-2 protease, and therefore inhibiting SARS-CoV-2 replication and proliferation. Searching the ZINC database for natural products with fortunellin as a bait, we have retrieved 16 natural compounds with similar molecular properties, presented in **Table S3 (Supplemental Material)**. Their absorption, distribution, metabolism, and excretion (ADME) characteristics were evaluated in the SwissADME site, and the results are shown in Supplemental Material **Table S4**. It derives that they are all water-soluble compounds, with limited enteric and skin absorption, but they are non-toxic, as they are not predicted to interact with the CYP drug metabolizing enzymes. Among them, apiin (ZINC3983878) and rhoifolin (ZINC3978800) have been previously studied for their biological effects. For further information please refer to the Supplemental Material section “Natural product Properties: the fortunellin – analogs”.

Our in-silico strategy identified a series of natural flavonoids – polyphenols (fortunellin, apiin, rhoifolin), which, in the form of drugs or dietary supplements, might be an effective strategy against the devastating SARS-CoV-2 infections, in humans.

**Acknowledgements**

We acknowledge PRACE for awarding us access to Joliot-Curie at GENCI@CEA (Irene), France, through the “PRACE support to mitigate impact of COVID-19 pandemic” call and the project “Epitope vaccines based on the dynamics of mutated SARS-CoV-2 proteins at all atom resolution”. We also acknowledge Greece and the European Union (European Social Fund- ESF) for funding through the Operational

Programme «Human Resources Development, Education and Lifelong Learning» in the context of the project "Strengthening Human Resources Research Potential via Doctorate Research" (MIS-5000432), implemented by the State Scholarships Foundation (IKY)» to AAP (PhD scholarship) and a Hellenic Foundation for Research and Innovation (H.F.R.I.) Grant to MK (# 3725).

**TOC Graph:**

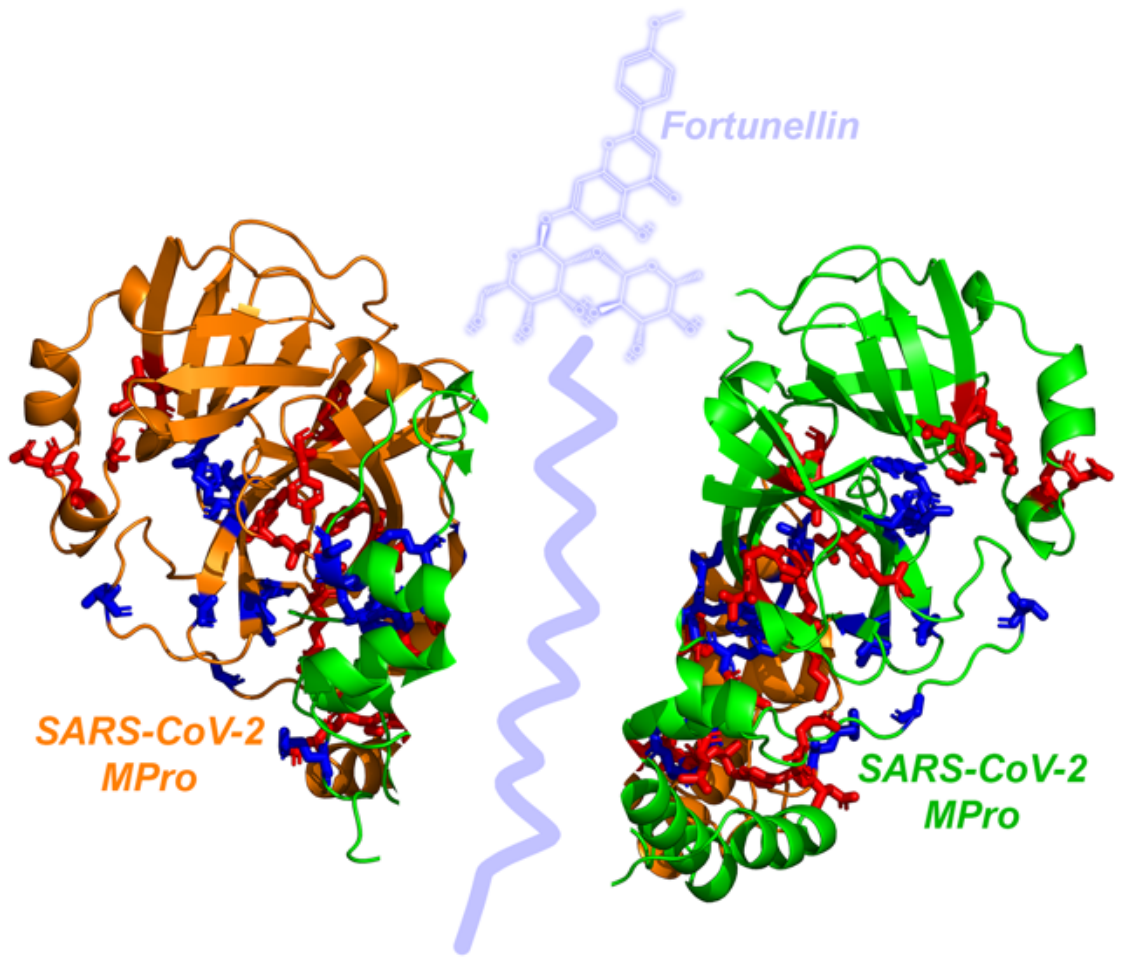

# Natural polyphenols inhibit the dimerization of the SARS-CoV-2 main protease: the case of fortunellin and its structural analogs

Athanasios A. Panagiotopoulos[*], Danai-Maria Kotzampasi[*], George Sourvinos[†,#], Marilena Kampa[*,#], Stergios Pirintsos[‡,&,#], Elias Castanas[*,#,‖], Vangelis Daskalakis[¶,‖]

* Laboratory of Experimental Endocrinology, University of Crete, School of Medicine, Heraklion, Greece

† Laboratory of Virology, University of Crete, School of Medicine, Heraklion, Greece

‡ Department of Biology, University of Crete, 71409 Heraklion, Greece

& Botanical Garden, University of Crete, Rethymnon, Greece

¶ Department of Chemical Engineering, Cyprus University of Technology, Limassol, Cyprus

# Nature Crete Pharmaceuticals, Heraklion, Greece

‖ Corresponding Authors

## Supplemental Material

# Contents

# Methods

## Molecular Docking

### SARS-CoV-2 main protease MPro and Ligand Preparation

The sequence of SARS-CoV-2 main protease MPro, in fasta format, was retrieved from the NCBI protein database (https://www.ncbi.nlm.nih.gov/protein/) and introduced to the Swiss Model Biospace (http://swissmodel.expasy.org/interactive).[1] The main SARS-CoV-2 protease MPro have many available crystal structures and the system, presented in Supplemental Table 1. Codes correspond to data stored in the Protein Data Bank (https://www.rcsb.org/).[2] In this work, we have used the 6YB7 unliganded crystal, analyzed with X-ray diffraction. Monomers were extracted, using a text editor.

Ligands were retrieved from the ZINC database (http://zinc.docking.org/),[3] in a canonical smiles format. Novel molecules were designed in ChemBioDraw (v12.0, Perkin Elmer, Boston, MA, free for Academic use from the University of Cambridge). Then, pdb or mol2 files of the ligands were created with the Open Babel program (http://openbabel.org).[4]

### Ligand-MPro docking

SARS-CoV-2 main protease (in pdb format) and ligand(s) (in mol2 or pdb formats) were uploaded to the GalaxyWebserver (http://galaxy.seoklab.org/) and a fully flexible docking (involving the receptor and the ligand) was performed.[5–10] The server uses an algorithm, based on the GalaxyDock2 docking,[11] which, after an automatic prediction of the ligand binding pocket, permits a full ligand/receptor flexibility during binding simulation. This step is followed by optimization and subsequent refinement, through a specific algorithm named GalaxyRefine,[6,12] which permits a protein-ligand structure refinement, by applying iterative side chain repacking and overall structure relaxation.[6,13] The best solution (usually denoted as “Model 1”) was retained. The corresponding pdb file (containing the MPro-ligand complex) was retrieved. The 3D structures of the liganded and unliganded MPro were compared, using the UCSF Chimera 1.11.2 program,[14] available from https://www.cgl.ucsf.edu/chimera/. The same program was used for comparisons of the retrieved solutions with available crystals

### Simulation of MPro dimerization

1) Dimerization of MPro: The active form of the MPro SARS-CoV-2 main protease, is a homodimer,[15] as a Cyclic-C2 (Global Symmetry) Homo-2-mer A2 (Global Stoichiometry).[16–19] Here, we used two different docking programs to simulate the active dimer and we compare the obtained solutions with the crystal 6YB7:

(1) We used the GalaxyWEB server, routine HOMOMER (http://galaxy.seoklab.org/cgi-bin/submit.cgi?type=HOMOMER) for the MPro protein oligomer structure prediction from a monomer sequence or structure (Ologomeric state = 2). The server carries out template-based modeling, *ab initio* docking or both, depending on the availability of proper oligomer templates, until 5 models are generated.[5]

(2) We used the Hex 8.0.8 program (http://hex.loria.fr/).[20,21] Hex 8.0.8, a specialized, locally executed, program, for protein-protein, or protein-nucleic acid interactions, based on spherical rotated protein complexes, taking into account both surface shape and electrostatic charge. Hex returns, through a graphical user interface, a set of >100 solutions, with the corresponding ΔG values. The input files for Hex 8.0.8 program are pdb files of MPro with or without ligand and the output file is a pdb file that contains the simulated MPro dimer. The 3D structures of the monomer and dimer structure of MPro from experimental crystal structures, GalaxyWEB server and Hex 8.0.8 program were compared, using the UCSF Chimera 1.11.2 program. The same program was also used to determine the dimerization interphase of the monomers.

## Molecular Dynamics computational protocol

### Model setup

The crystal structures of SARS-CoV-2 main protease MPro, or 3CL-PRO (pdb codes 6LU7[22] and 6YB7) were used as initial coordinates to build the models. The pdb structures refer to one monomer without and with inhibitor, respectively. For consistency, the inhibitor was removed from the crystal structure of MPro (6LU7).[22] Our choice of coordinates was based on the completeness of the resolved MPro sequence and the quality of chains (at least 90%). To build the MPro dimer, we structurally aligned either two 6lu7, or two 6yb7 monomers on a reference dimer.[23] The protonation states of titratable residues were simulated at neutral pH, thus all Glu, and Asp residues were left deprotonated, except Glu-290 which was protonated, in accordance also with the PDB2PQR (propka 3.0 method, pH 7.3) predictions.[24] His-41, His-163, His-172 and His-246 were protonated only at the Nε site. The rest of His residues were protonated only at the Nδ sites, to maintain the hydrogen bonding network within the crystal structures. All crystallographic water molecules are retained within each crystal structure. Four samples were, thus prepared, in a consistent way (two monomers, two dimers). The all-atom models, as defined previously, were embedded in triclinic boxes of around 7.2nm x 11.2nm x 8.0nm (monomer), or 12.3nm x 12.3 nm x 12.3 nm (dimer) in the x, y and z dimensions, respectively. Up to around 57000 TIP3P water molecules[25] were used to hydrate each protein. Ion ($K^+$, $Cl^-$) concentration was set at the value of 150 mM to mimic the physiological salt content for the monomer or dimer models, in addition to zero, or excess concentrations of KCl, NaCl, $CaCl_2$ at 0, 150, 300, 400 and 500 mM only for the monomer models. The various anionic strengths were only employed to indirectly 'enhance' the sampled conformational space of the MPro. A surplus of $K^+$ was also added to neutralize the protein charges in each sample, resulting in simulation unit boxes of 62400 (monomer), or 181800 (dimer) atoms. The Amber03[26] protein force field was used for the residues and ions. The Amber03 parameters for the natural products were derived based on ACPYPE algorithm.[27]

### Equilibration-Production Molecular Dynamics setup

The equilibration-relaxation for the all-atom systems is employed based on a published protocol for water-soluble proteins.[28] This contains a steepest descend energy minimization with a tolerance of 0.5 kJ $mol^{-1}$ for 1000 steps, and a sequence of isothermal (nVT), isothermal-isobaric (nPT) runs with the gradual relaxation of the constraints on protein heavy atoms (from $10^4$ in steps1-2 to $10^3$ kJ $mol^{-1}$ $nm^{-2}$ in step-4) and Cα atoms (from $10^3$ in step-5, to $10^2$ in step-6, 10 in step-7, 1 in step-8 and 0 kJ $mol^{-1}$ $nm^{-2}$ in step-9) for around 30 ns, with a time step of 1.0 fs (steps 1-4) and 2.0 fs (steps 5-9). In detail: (**step-1**) Constant density and temperature (nVT) Brownian dynamics (BD) at 100 K for 50 ps that employs the Berendsen thermostat,[29] with a temperature coupling constant at 1.0 fs. (**steps 2-3**) Two short constant density (nVT) and constant pressure (nPT) runs for 100 ps each, with a weak coupling Berendsen thermostat and barostat[29] at 100 K employing time coupling constants of 0.1 ps for the temperature and isotropic 50.0 ps coupling for the pressure with a compressibility of $4.6x10^{-5}$. (**step-4**) Heating from 100 to 250 K in a constant density ensemble (nVT) for 3 ns employing the v-rescale thermostat,[30] with a time coupling constant of 0.1 ps. (**step-5**) Heating from 250 to 310K in a constant pressure ensemble (nPT) for 2 ns, employing the v-rescale thermostat[30] and Berendsen barostat,[29] with time coupling constants of 0.1 ps for the temperature and 2.0 ps for the pressure, removing also all but the Cα-atom protein position restraints. (**step-6**) Equilibration at 310K (0.1 ps temperature coupling constant) for 5 ns (nPT, 1 atm, 2.0 ps coupling constant for pressure. (**steps 7-8**) Equilibration at 310K (0.5 ps temperature coupling constant) for 5 ns (nPT, 1 atm, 2.0 ps coupling constant for pressure). (**step-9**) Equilibration at 310K (0.5 ps temperature coupling constant) for 10 ns (nPT, 1 atm, 2.0 ps coupling constant for pressure). The barostats – thermostats employed for steps 6-9 were the same as in the production trajectories that follow.

For the production all-atom classical Molecular Dynamics (MD), the Newton's equations of motion are integrated with a time step of 2.0 fs at 310K. All production simulations are run with the leap-frog integrator in GROMACS 2020[31] for 3.0 μs each. They were performed at the constant pressure nPT ensemble, with isotropic coupling (compressibility at $4.5x10^{-5}$) employing the v-rescale thermostat[30] (310K, temperature coupling constant 0.5) and the *Parrinello-Rahman* barostat[32] (1 atm, pressure coupling constant 2.0). Details for parameters can be found in earlier work.[28] The first 500 ns were considered further equilibration from each independent trajectory per sample, and were disregarded in the analysis. Van der Waals interactions were smoothly switched to zero between 1.0-1.2 nm with the VERLET cut-off scheme. Electrostatic interactions were truncated at 1.2 nm (short-range) and long-range contributions were computed within the PME approximation.[33,34] Hydrogen bond lengths were constrained employing the LINCS algorithm.[35]

## Markov State Model Analysis

We obtained a series of MD equilibrium trajectories of the MPro monomers (60μs) of SARS-CoV-2 main protease under different salts/ ionic strengths, without inhibitors. These should have explored a major part of the MPro conformational phase space. We combined the all-atom MD simulations with Markov state model (MSM) theory[36–38] in order to enable the extraction of long-time-scale individual monomer dynamics from rather short-time-scale MD trajectories of different states. The application and accuracy of the powerful MSM theory has been presented in many cases also by experiments that include protein−protein, or protein-drug binding kinetics, as well as protein folding rates and protein dynamics.[39–42] Our objective was to approximate the slow dynamics in a statistically efficient manner. Thus, a lower dimensional representation of our simulation data was necessary. In order to reduce the dimensionality of our feature space, we employed the time-structure independent components analysis (tICA) which yields a representation of our molecular simulation data with a reduced dimensionality and can greatly facilitate the decomposition of our system into the discrete Markovian states necessary for MSM estimation. The conformations of the system were projected on these slowest modes as defined by the tICA method, then the trajectory frames were clustered into 100 cluster-centers (microstates) by k-means clustering, as implemented in PyEMMA.[43] Conformational changes of a system can be simulated as a Markov chain, if the transitions between the different conformations are sampled at long enough time intervals so that each transition is Markovian. This means that a transition from one conformation to another is independent of the previous transitions. Therefore, an MSM is a memoryless model. The uncertainty bounds were computed using a Bayesian scheme.[44,45] We found that the slowest implied timescales converge quickly and are constant within a 95% confidence interval for lag times above 50ns. The validation procedure is a standard approach in the MSM field., a lag time of 50 ns was selected for Bayesian model construction, and the resulting models were validated by the Chapman-Kolmogorov (CK) test. Subsequently, the resulting MSMs were further coarse grained into a smaller number of three metastable states or microstates, using PCCA++ as implemented in PyEMMA.[43] The optimum number of microstates (three) was proposed based on the VAMP2-score.[46] Both the convergence of the implied timescales, as well as the CK test confirm the validity and convergence of the MSM. The CK test indicates that predictions from the built MSM agree well with MSMs estimated with longer lag times. Thus, the model can describe well the long-time-scale behavior of our system within error. The tICA method identified the torsional angles of the following MPro residues: 3, 4, 5, 6, 84, 135, 141, 164, 167, 171, 175, 178, 179, 180, 190, 195, 217, 284, 285, 286, 290, 291, 300, and 301 as the most important features, by setting a series of thresholds for the coefficients in the tICA vectors. At first, we set a threshold of 0.09. We continued by setting a threshold of 0.04 for the coefficients in the tICA vectors of the filtered data and afterwards a threshold of 0.085. Finally, we set a threshold of 0.075 and thus we concluded in the previously referred MPro residues. For the selection of these thresholds we checked for different thresholds the VAMP2-score and the states projected onto the first two independent components. We report the exact functions of the first two tICA components, as a linear combination of cosine/ sine functions of the associated torsionals:

CV1 = (0.22245479085963174)*COS(PHE3PHI)+(-0.16733307515910162)*SIN(PHE3PHI)+(0.1432275835003262)*COS(PHE3PSI)+(0.2008868998696375)*SIN(PHE3PSI)+(0.19936685751147043)*COS(ARG4PHI)+(-0.09327647115497929)*SIN(ARG4PHI)+(-0.0930358300031665)*COS(ARG4PSI)+(-0.08538905470780045)*SIN(ARG4PSI)+(0.10166329580502668)*COS(LYS5PHI)+(-0.09568173783678022)*SIN(LYS5PHI)+(-0.0928774945984135)*COS(LYS5PSI)+(-

0.15790698786616095)*SIN(LYS5PSI)+(-
0.0956189898931311)*COS(MET6PHI)+(0.025245999648597445)*SIN(MET6PHI)+(0.0408771827310
96175)*COS(MET6PSI)+(-
0.0006973055782739627)*SIN(MET6PSI)+(0.12585421365826116)*COS(ASN84PHI)+(0.1980318831
213358)*SIN(ASN84PHI)+(0.08785335907805712)*COS(ASN84PSI)+(-
0.1678608618927701)*SIN(ASN84PSI)+(0.3615169358035203)*COS(THR135PHI)+(0.2355840179344
413)*SIN(THR135PHI)+(0.06937744998780684)*COS(THR135PSI)+(-
0.08624991337144639)*SIN(THR135PSI)+(0.06993246492427758)*COS(LEU141PHI)+(-
0.023980684220480688)*SIN(LEU141PHI)+(-0.0028980923592885344)*COS(LEU141PSI)+(-
0.11124570694911744)*SIN(LEU141PSI)+(0.12367941924553509)*COS(HIS164PHI)+(0.1100289626
0080477)*SIN(HIS164PHI)+(0.1736049601096124)*COS(HIS164PSI)+(-
0.10071229902364524)*SIN(HIS164PSI)+(0.04589771207808335)*COS(LEU167PHI)+(0.0319038117
3082644)*SIN(LEU167PHI)+(0.010408062573537607)*COS(LEU167PSI)+(-
0.001452825004362783)*SIN(LEU167PSI)+(-
0.035449253177574046)*COS(VAL171PHI)+(0.0397831323983392)*SIN(VAL171PHI)+(-
0.01487903934140085)*COS(VAL171PSI)+(-
0.034102380825128384)*SIN(VAL171PSI)+(0.030847079909921624)*COS(THR175PHI)+(-
0.020040648643118112)*SIN(THR175PHI)+(0.12177024626152966)*COS(THR175PSI)+(-
0.13229839249150296)*SIN(THR175PSI)+(0.11507939509465313)*COS(GLU178PHI)+(0.1013714392
35159)*SIN(GLU178PHI)+(-
0.09137875539957069)*COS(GLU178PSI)+(0.11214339094322172)*SIN(GLU178PSI)+(-
0.09294920607798879)*COS(GLY179PHI)+(0.28997810094722837)*SIN(GLY179PHI)+(-
0.27573006050671983)*COS(GLY179PSI)+(-
0.09673320535071989)*SIN(GLY179PSI)+(0.010926172367439396)*COS(ASN180PHI)+(-
0.24668920154301732)*SIN(ASN180PHI)+(-
0.14912645633298552)*COS(ASN180PSI)+(0.3816102785675176)*SIN(ASN180PSI)+(-
0.10964069317986454)*COS(THR190PHI)+(-0.009927774218497826)*SIN(THR190PHI)+(-
0.16097973924205633)*COS(THR190PSI)+(-
0.16639818413751062)*SIN(THR190PSI)+(0.08017482811083185)*COS(GLY195PHI)+(-
0.14979887114706295)*SIN(GLY195PHI)+(-0.050958292897186695)*COS(GLY195PSI)+(-
0.044354361874655926)*SIN(GLY195PSI)+(0.09987054135337944)*COS(ARG217PHI)+(-
0.12648381947691048)*SIN(ARG217PHI)+(0.08158948319042672)*COS(ARG217PSI)+(-
0.00021426690460247595)*SIN(ARG217PSI)+(0.21642336166830292)*COS(SER284PHI)+(-
0.30155833144467264)*SIN(SER284PHI)+(0.8266406109913902)*COS(SER284PSI)+(0.88748800878
14281)*SIN(SER284PSI)+(0.6649555215066218)*COS(ALA285PHI)+(0.44940445619426095)*SIN(ALA
285PHI)+(-0.7837894441282054)*COS(ALA285PSI)+(-0.7618813787744232)*SIN(ALA285PSI)+(-
0.41060021549010234)*COS(LEU286PHI)+(-
0.10014769031761044)*SIN(LEU286PHI)+(0.19732367941536325)*COS(LEU286PSI)+(0.2864165371
4402343)*SIN(LEU286PSI)+(0.2011752565537491)*COS(GLU290PHI)+(0.08698680780975648)*SIN(
GLU290PHI)+(-
0.00570069571470415)*COS(GLU290PSI)+(0.06637101699247436)*SIN(GLU290PSI)+(0.2298963479
5972827)*COS(PHE291PHI)+(-0.17343642924439104)*SIN(PHE291PHI)+(-
0.11571250082711405)*COS(PHE291PSI)+(0.02324511232530362)*SIN(PHE291PSI)+(-
0.047478679669659275)*COS(CYS300PHI)+(0.05578968284901458)*SIN(CYS300PHI)+(-
0.1683363737052317)*COS(CYS300PSI)+(-0.2973909304454135)*SIN(CYS300PSI)+(-

0.1038665536799251)*COS(SER301PHI)+(-0.019617745122528882)*SIN(SER301PHI)+(-0.06934626053471138)*COS(SER301PSI)+(-0.10587162831562456)*SIN(SER301PSI)

CV2 = (-0.2884241825864551)*COS(PHE3PHI)+(0.1818110423100234)*SIN(PHE3PHI)+(-0.4677450443427827)*COS(PHE3PSI)+(-0.5483897439796918)*SIN(PHE3PSI)+(-0.3553107265200588)*COS(ARG4PHI)+(-0.22394962064864046)*SIN(ARG4PHI)+(-0.270179755725878)*COS(ARG4PSI)+(-0.3829402238055943)*SIN(ARG4PSI)+(-0.2042142888426065)*COS(LYS5PHI)+(-0.1669302788926349)*SIN(LYS5PHI)+(0.141973243784881)*COS(LYS5PSI)+(0.1421901057216299)*SIN(LYS5PSI)+(0.0846185601658726)*COS(MET6PHI)+(0.20450776328031908)*SIN(MET6PHI)+(-0.03312223507339774)*COS(MET6PSI)+(-0.08043115352475075)*SIN(MET6PSI)+(-0.027921418396744165)*COS(ASN84PHI)+(-0.08339140826693826)*SIN(ASN84PHI)+(-0.38318457142235807)*COS(ASN84PSI)+(-0.02014396174778721)*SIN(ASN84PSI)+(-0.261602759573732)*COS(THR135PHI)+(-0.22169542813755444)*SIN(THR135PHI)+(-0.087566095559111)*COS(THR135PSI)+(0.11762076691089288)*SIN(THR135PSI)+(-0.3469970284275915)*COS(LEU141PHI)+(0.34989926232122653)*SIN(LEU141PHI)+(-0.3353014777111922)*COS(LEU141PSI)+(-0.1099813590134305)*SIN(LEU141PSI)+(-0.030401089904733867)*COS(HIS164PHI)+(-0.12811671967132546)*SIN(HIS164PHI)+(-0.13436881803853637)*COS(HIS164PSI)+(0.17139820505530204)*SIN(HIS164PSI)+(-0.12200117274899505)*COS(LEU167PHI)+(-0.06149505441676782)*SIN(LEU167PHI)+(-0.06260851492073659)*COS(LEU167PSI)+(0.06084287820758976)*SIN(LEU167PSI)+(-0.006861594758294066)*COS(VAL171PHI)+(0.012827540881062702)*SIN(VAL171PHI)+(0.012464087205759892)*COS(VAL171PSI)+(-0.02648078463371077)*SIN(VAL171PSI)+(0.13636122184567023)*COS(THR175PHI)+(-0.16662872008089358)*SIN(THR175PHI)+(0.005026593046059509)*COS(THR175PSI)+(-0.020823512651369957)*SIN(THR175PSI)+(-0.15569457005305837)*COS(GLU178PHI)+(-0.12910762053786512)*SIN(GLU178PHI)+(0.195962425618426)*COS(GLU178PSI)+(-0.1674075865238855)*SIN(GLU178PSI)+(-0.05538839797914528)*COS(GLY179PHI)+(-0.2984068151828335)*SIN(GLY179PHI)+(0.4196490331550077)*COS(GLY179PSI)+(0.2097945401494899)*SIN(GLY179PSI)+(-0.021574332218825033)*COS(ASN180PHI)+(0.3273623222055997)*SIN(ASN180PHI)+(0.08717870414956289)*COS(ASN180PSI)+(-0.47359288885540496)*SIN(ASN180PSI)+(-0.04538558390983336)*COS(THR190PHI)+(-0.008683270082020167)*SIN(THR190PHI)+(0.271129910670838)*COS(THR190PSI)+(0.27549665651127636)*SIN(THR190PSI)+(0.0780842690641956)*COS(GLY195PHI)+(0.044233409296374225)*SIN(GLY195PHI)+(0.13353598315320972)*COS(GLY195PSI)+(0.04465259963400809)*SIN(GLY195PSI)+(0.027701779053662518)*COS(ARG217PHI)+(0.03397617632773808)*SIN(ARG217PHI)+(-0.055406779213062196)*COS(ARG217PSI)+(-0.15249136969809435)*SIN(ARG217PSI)+(0.15917442405868654)*COS(SER284PHI)+(0.07317395309230672)*SIN(SER284PHI)+(0.050094307429749026)*COS(SER284PSI)+(0.19617816860701598)*SIN(SER284PSI)+(0.1826732916695155)*COS(ALA285PHI)+(-0.035338601991516574)*SIN(ALA285PHI)+(-0.19924762948366187)*COS(ALA285PSI)+(-0.1564466277523662)*SIN(ALA285PSI)+(-0.0443686239963879)*COS(LEU286PHI)+(-0.10931048704118541)*SIN(LEU286PHI)+(0.45176938273388506)*COS(LEU286PSI)+(0.4751207479

763982)*SIN(LEU286PSI)+(0.1954750353189944)*COS(GLU290PHI)+(0.28698450313441914)*SIN(GLU290PHI)+(0.1178561297010931)*COS(GLU290PSI)+(-0.11293348016651641)*SIN(GLU290PSI)+(-0.23109368829696822)*COS(PHE291PHI)+(0.2905179445957612)*SIN(PHE291PHI)+(0.2414087396867373)*COS(PHE291PSI)+(-0.19323071296694158)*SIN(PHE291PSI)+(-0.06723387446469688)*COS(CYS300PHI)+(0.08472356860436044)*SIN(CYS300PHI)+(0.11286902793988102)*COS(CYS300PSI)+(0.33756361790231243)*SIN(CYS300PSI)+(0.07472554678618155)*COS(SER301PHI)+(-0.04927948273422832)*SIN(SER301PHI)+(0.13667888455911714)*COS(SER301PSI)+(0.1947392228687667)*SIN(SER301PSI)

The residues that seem to affect the MPro-fortunellin dynamics are derived based on the same MSM/ tICA protocol described before in the absence of the inhibitor. These include: 44, 48, 53, 82, 83, 84, 111, 112, 118, 137, 138, 139, 141, 159, 182, 238, 239, 240, 286, 287, 288, 289, 291. At first, we set a threshold of 0.075. We continued by setting a threshold of 0.12 for the coefficients in the tICA vectors of the filtered data, a threshold of 0.18, a threshold of 0.065 and afterwards a threshold of 0.135. Finally, we set a threshold of 0.125 and thus we concluded in the previously referred residues. However, the MPro states in the presence of fortunellin, reported in Figure 1E of the main manuscript, are derived based on the projection of the Mpro-fortunellin trajectory data on the aforementioned phase space of the CV1, CV2 functions.

## Enhanced Molecular Dynamics Sampling

To enhance the conformational sampling on the MPro dimer we employed the parallel tempering metadynamics in the well-tempered ensemble (PTmetaD-WTE) method.[47–50] Nine replicas per sample were run at 310, 320, 330, 341, 352, and 363K, 375K, 387K and 400K in which only the potential energy (PE) was initially biased (bias factor 120) to achieving large fluctuations in PE and replica overlaps. Replicas were allowed to exchange every 1000 steps for 0.2μs each, which gave an exchange probability around 20% in the WTE. The obtained bias was saved and used for the subsequent PTmetaD production runs for another 0.5μs per sample/ replica. Nine replicas were again considered at the same temperatures. Including the equilibration time at reach replica, a cumulative simulation time of 10μs was achieved. An exchange was attempted every 1000 steps, that gave an exchange probability between replicas at around 20%, consistent with the large sample sizes. The Collective Variables (CVs) chosen for the PTmetaD runs were the first two tICA vectors presented above (CV1/ CV2). A combination of the GROMACS 2020/ PLUMED 2.5[51] engines was employed. A bias factor of 25 at the well-tempered ensemble, along with Gaussians of 1.2 kJ/mol initial height, and sigma values (width) of 0.25 in the CV space, deposited every 2 ps, was employed. The grid space for both CVs is defined between -4 to 4 at a resolution of 0.05. Four different PTmetaD-WTE runs were performed for the pdb 6yb7/ 6lub7 -based dimers at 150mM KCl without inhibitor.

## Natural product properties: the fortunellin – analogs

Apiin (apigenin-7-apioglucoside), is a natural flavonoid, a diglycoside of the flavone apigenin, isolated from leaves of *Apium graveolens* var. *dulce* (celery) and *Petroselinum crispum* (parsley). Apiin extracted from celery exhibited anti-inflammatory properties, as apiin showed strong inhibitory activity on inducible nitric oxide synthase (iNOS) expression and nitrite (NO) production when added before LPS stimulation of J774.A1 cells.[52] In mice models, apiin had a remarkable scavenging activity on maleic dialdehyde (MDA) and lipofuscin (LPF), promoted total antioxidant capacity (TAOC) and significantly enhanced the activities of superoxide dismutase (SOD), glutathione peroxidase (GSH-Px) and catalase (CAT),[53] by exerting a radical scavenging activity greater than that of absorbic acid[54] and Vitamin E.[53] Apiin also showed a marked antihypertensive effect in experimental pulmonary hypertension in dogs [55] and anti-influenza virus activity *in vitro* through inhibition of neuraminidase (NA).[56] Besides, the role of apiin cardiovascular activity as antiarrhythmic and anti-ischemic agent has also been reported.[57] In view of our results, apiin might therefore be a strong drug candidate, as it inhibits SARS-CoV-2 virus and tackling also COVID-19 major disease symptoms.

Rhoifolin (apigenin 7-O-neohesperidoside), is a neohesperidoside, a dihydroxyflavone and a glycosyloxyflavone, was first isolated from plant *Rhus succedanea* (Sumac or wax tree, originating from Asia, but also found in Australia and New Zealand).[58] Resently, rhoifolin was found to efficiently block the enzymatic activity of SARS-CoV 3CL-Pro,[59] with a methodology similar to that used in the present study. In addition, rhoifollin was reported to inhibit CVB3 infection, a primary cause of viral myocarditis in humans. In addition, it was found to decrease inflammation, by significantly decreasing prostaglandin E2 and the release of pro-inflammatory cytokines (TNF-α, IL-1β, and IL-6).[60,61] Rhoifolin isolated from *Citrus grandis* leaves was beneficial in metabolic diseases, including type II diabetes, by enhancing adiponectin secretion, tyrosine phosphorylation of insulin receptor-β and GLUT4 translocation.[62] Rhoifolin also caused a decrease of mean aortic pressure, of the arterial and pulmonary capillary pressure and of heart rate in the dog.[63] Moreover, previous study has demonstrated that rhoifolin can have an inhibitory effect on angiotensin-converting enzyme (ACE) activity, which plays a key role in the regulation of arterial blood pressure.[63]

## Supplemental Table S1

*PDB codes of 3CL-PRO crystals, deposited in the PDB databank, accessed at July 3, 2020.*

5r7y, 5r7z, 5r80, 5r81, 5r82, 5r83, 5r84, 5r8t, 5re4, 5re5, 5re6, 5re7, 5re8, 5re9, 5rea, 5reb, 5rec, 5red, 5ree, 5ref, 5reg, 5reh, 5rei, 5rej, 5rek, 5rel, 5rem, 5ren, 5reo, 5rep, 5rer, 5res, 5ret, 5reu, 5rev, 5rew, 5rex, 5rey, 5rez, 5rf0, 5rf1, 5rf2, 5rf3, 5rf4, 5rf5, 5rf6, 5rf7, 5rf8, 5rf9, 5rfa, 5rfb, 5rfc, 5rfd, 5rfe, 5rff, 5rfg, 5rfh, 5rfi, 5rfj, 5rfk, 5rfl, 5rfm, 5rfn, 5rfo, 5rfp, 5rfq, 5rfr, 5rfs, 5rft, 5rfu, 5rfv, 5rfw, 5rfx, 5rfy, 5rfz, 5rg0, 5rg1, 5rg2, 5rg3, 5rgg, 5rgh, 5rgi, 5rgj, 5rgk, 5rgl, 5rgm, 5rgn, 5rgo, 5rgp, 5rgq, 5rgr, 5rgs, 5rgt, 5rgu, 5rgv, 5rgw, 5rgx, 5rgy, 5rgz, 5rh0, 5rh1, 5rh2, 5rh3, 5rh4, 5rh5, 5rh6, 5rh7, 5rh8, 5rh9, 5rha, 5rhb, 5rhc, 5rhd, 5rhe, 5rhf, 6lu7, 6lze, 6m03, 6m0k, 6m2n, 6m2q, 6w63, 6wnp, 6wqf, 6wtj, 6wtk, 6wtm, 6wtt, 6xa4, 6xb0, 6xb1, 6xb2, 6xbg, 6xbh, 6xbi, 6xch, 6y2e, 6y2f, 6y2g, 6y84, 6yb7, 6ynq, 6yt8, 6yvf, 6yz6, 6z2e, 7bqy, 7bro, 7brp, 7brr, 7buy, 7c8r, 7c8t

## Supplemental Table S2

Table presents the global and local RMSD values, at the dimerization domain) of 3CL-Pro monomer, at different time-frames, as determined by molecular dynamics. The interaction of fortunellin in each pose is also shown, as changes of the Gibbs free energy changes (ΔG).

| Model | Time (μs) | RMSD Total | RMSD Local | ΔG (kcal/mol) |
|---|---|---|---|---|
| 0 | 0 | 0.000 | 0.000 | -13.936 |
| 1 | 3 | 1.637 | 1.235 | -11.199 |
| 2 | 6 | 2.248 | 3.290 | -12.144 |
| 3 | 9 | 1.954 | 1.515 | -10.704 |
| 4 | 12 | 1.421 | 1.292 | -14.213 |
| 5 | 15 | 1.756 | 1.085 | -11.611 |
| 6 | 18 | 1.706 | 0.998 | -11.987 |
| 7 | 21 | 2.292 | 2.608 | -14.658 |
| 8 | 24 | 2.694 | 2.757 | -13.699 |
| 9 | 27 | 1.845 | 1.230 | -13.563 |
| 10 | 30 | 1.821 | 1.268 | -11.468 |
| 11 | 33 | 2.031 | 1.344 | -12.760 |
| 12 | 36 | 1.924 | 1.604 | -12.518 |
| 13 | 39 | 1.888 | 1.269 | -11.579 |
| 14 | 42 | 2.544 | 2.686 | -12.165 |
| 15 | 45 | 1.897 | 1.463 | -11.507 |
| 16 | 48 | 2.122 | 1.883 | -12.423 |
| 17 | 51 | 1.965 | 1.420 | -11.478 |
| 18 | 54 | 1.924 | 1.534 | -11.997 |
| 19 | 57 | 1.915 | 1.263 | -12.070 |
| 20 | 60 | 1.874 | 1.385 | -12.660 |

## Supplemental Table S3

ZINC number, formula, SMILES and binding affinity for 3CL-Pro (as expressed by Gibb's free energy, ΔG, change in kcal/mol) of ther sixteen (16) compounds from the ZINC database for natural products.

| Compound | 2D Structure | Smiles | ΔG (Kcal/mol) |
|---|---|---|---|
| ZINC4349204 (Fortunellin) | 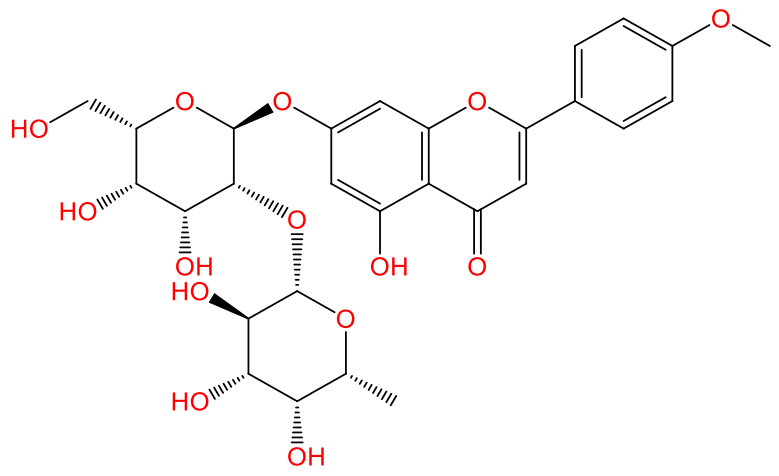 | COc1ccc(-c2cc(=O)c3c(O)cc(O[C@@H]4O[C@@H](CO)[C@@H](O)[C@@H](O)[C@H]4O[C@@H]4O[C@H](C)[C@H](O)[C@H](O)[C@H]4O)cc3o2)cc1 | -13.936 |
| ZINC4349029 | 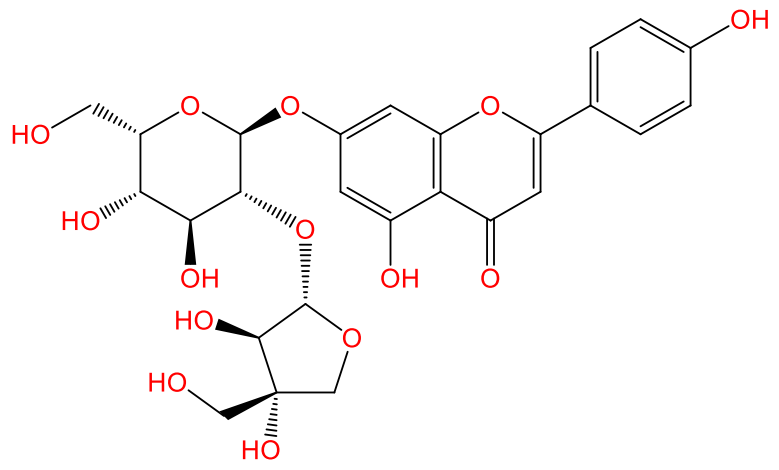 | O=c1cc(-c2ccc(O)cc2)oc2cc(O[C@@H]3O[C@@H](CO)[C@@H](O)[C@H](O)[C@H]3O[C@@H]3OC[C@@](O)(CO)[C@H]3O)cc(O)c12 | -14.241 |
| ZINC4349031 | 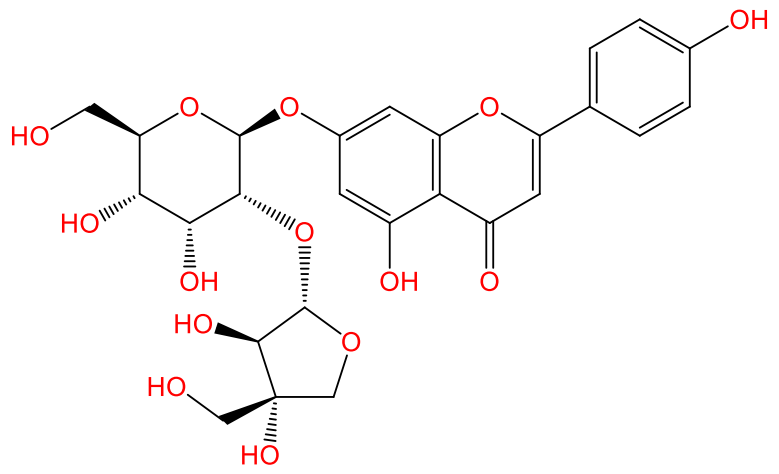 | O=c1cc(-c2ccc(O)cc2)oc2cc(O[C@@H]3O[C@H](CO)[C@@H](O)[C@@H](O)[C@H]3O[C@@H]3OC[C@@](O)(CO)[C@H]3O)cc(O)c12 | -12.352 |
| ZINC4349034 | 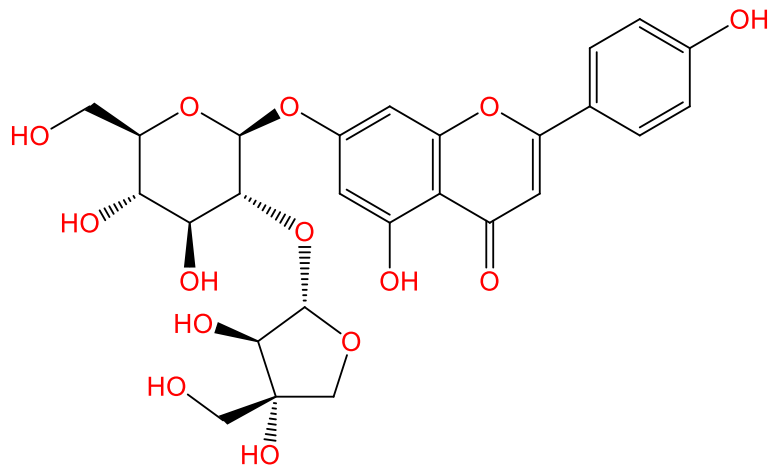 | O=c1cc(-c2ccc(O)cc2)oc2cc(O[C@@H]3O[C@H](CO)[C@@H](O)[C@H](O)[C@H]3O[C@@H]3OC[C@@](O)(CO)[C@H]3O)cc(O)c12 | -12.451 |
| ZINC3983878 (Apiin) | 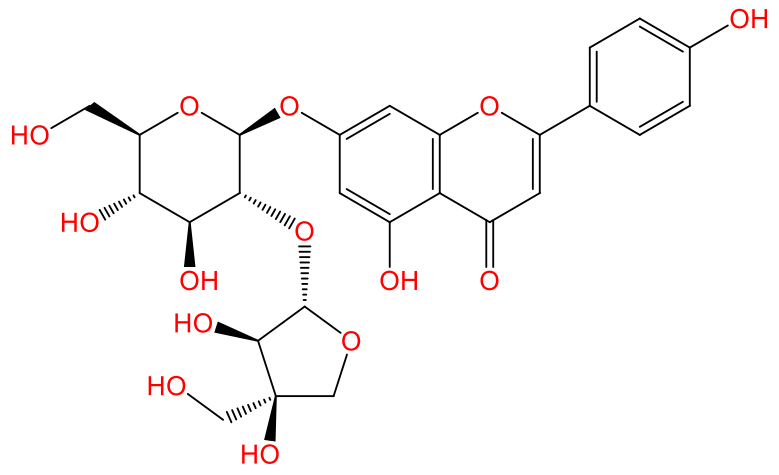 | O=c1cc(-c2ccc(O)cc2)oc2cc(O[C@@H]3O[C@H](CO)[C@@H](O)[C@H](O)[C@H]3O[C@@H]3OC[C@](O)(CO)[C@H]3O)cc(O)c12 | -12.604 |
| ZINC4349207 | 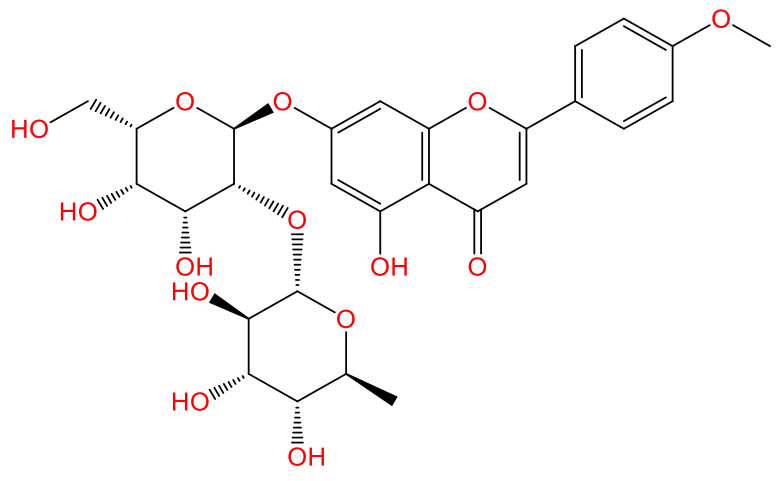 | COc1ccc(-c2cc(=O)c3c(O)cc(O[C@@H]4O[C@@H](CO)[C@@H](O)[C@@H](O)[C@H]4O[C@@H]4O[C@@H](C)[C@H](O)[C@H](O)[C@H]4O)cc3o2)cc1 | -11.539 |
| ZINC4349211 | 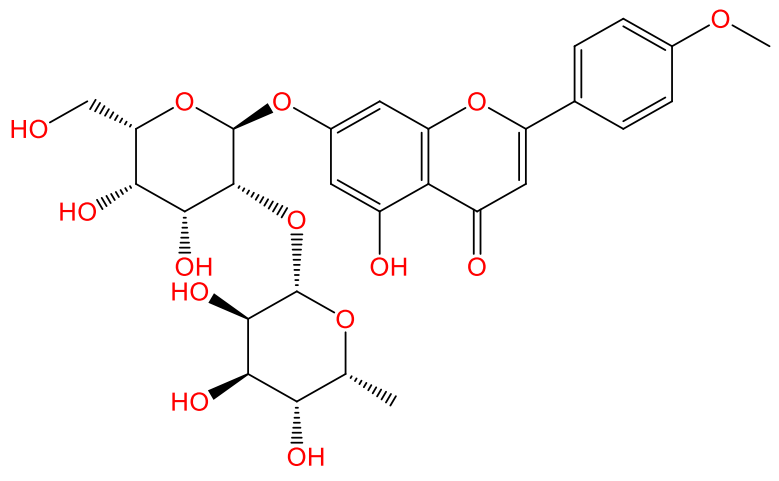 | COc1ccc(-c2cc(=O)c3c(O)cc(O[C@@H]4O[C@@H](CO)[C@@H](O)[C@@H](O)[C@H]4O[C@@H]4O[C@H](C)[C@H](O)[C@@H](O)[C@H]4O)cc3o2)cc1 | -11.545 |
| ZINC4349214 | 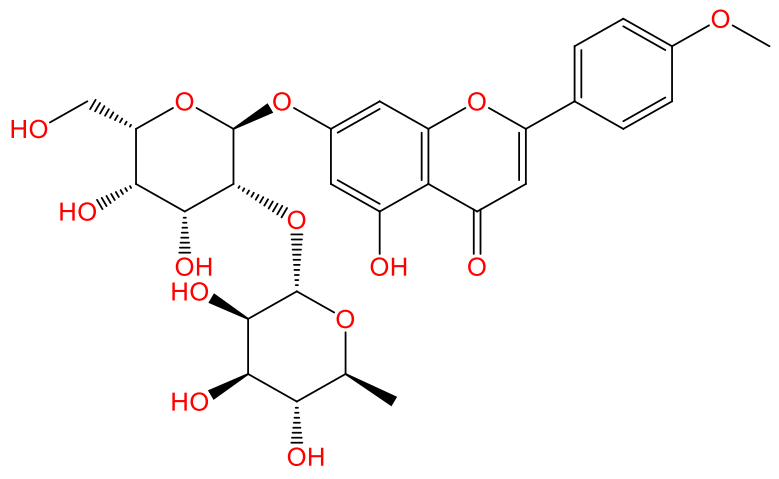 | COc1ccc(-c2cc(=O)c3c(O)cc(O[C@@H]4O[C@@H](CO)[C@@H](O)[C@@H](O)[C@H]4O[C@@H]4O[C@@H](C)[C@H](O)[C@@H](O)[C@H]4O)cc3o2)cc1 | -12.150 |
| ZINC4349623 | 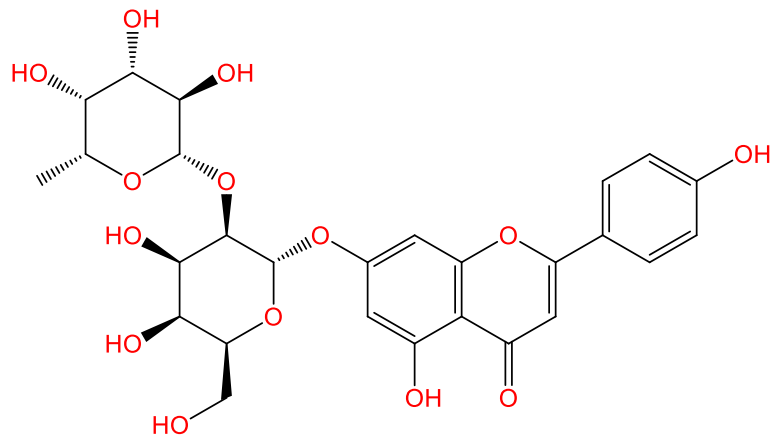 | C[C@H]1O[C@@H](O[C@H]2[C@H](Oc3cc(O)c4c(=O)cc(-c5ccc(O)cc5)oc4c3)O[C@@H](CO)[C@@H](O)[C@H]2O)[C@H](O)[C@@H](O)[C@H]1O | -12.970 |
| ZINC4349627 | 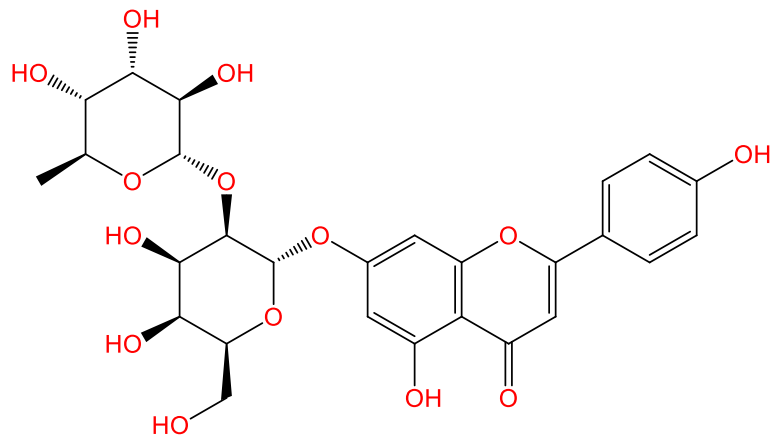 | C[C@@H]1O[C@@H](O[C@H]2[C@H](Oc3cc(O)c4c(=O)cc(-c5ccc(O)cc5)oc4c3)O[C@@H](CO)[C@@H](O)[C@H]2O)[C@H](O)[C@@H](O)[C@H]1O | -13.200 |
| ZINC4349630 | 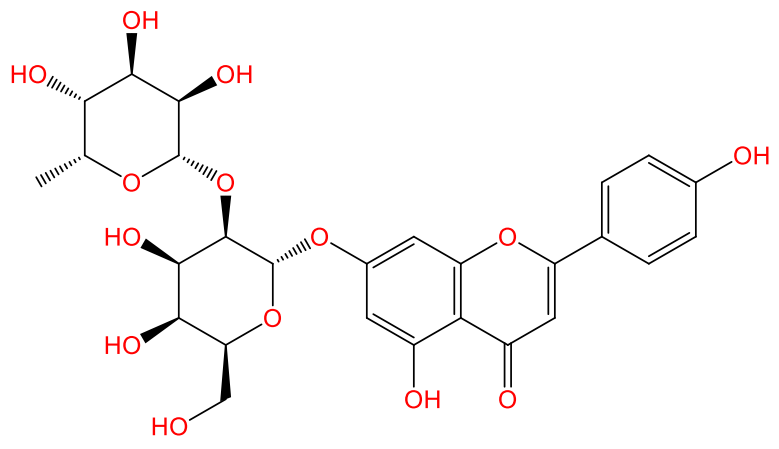 | C[C@H]1O[C@@H](O[C@H]2[C@H](Oc3cc(O)c4c(=O)cc(-c5ccc(O)cc5)oc4c3)O[C@@H](CO)[C@@H] | -12.229 |
| | | (O)[C@H]2O)[C@H](O)[C@H](O)[C@H]1O | |
| ZINC4349633 | 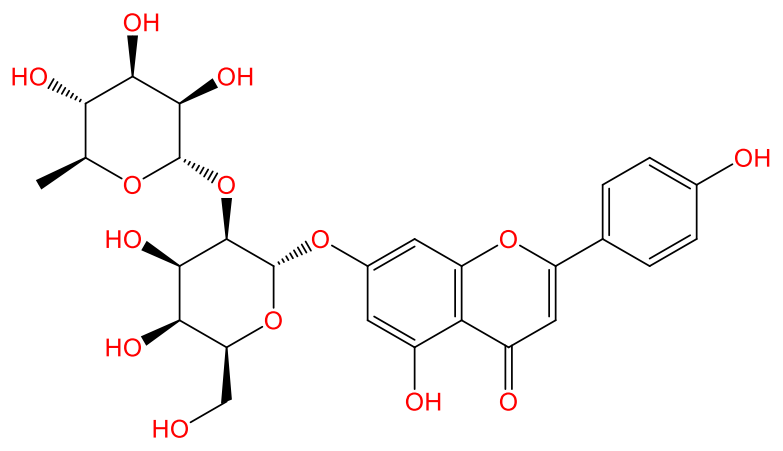 | C[C@@H]1O[C@@H](O[C@H]2[C@H](Oc3cc(O)c4c(=O)cc(-c5ccc(O)cc5)oc4c3)O[C@@H](CO)[C@@H](O)[C@H]2O)[C@H](O)[C@H](O)[C@H]1O | -12.937 |
| ZINC4534057 | 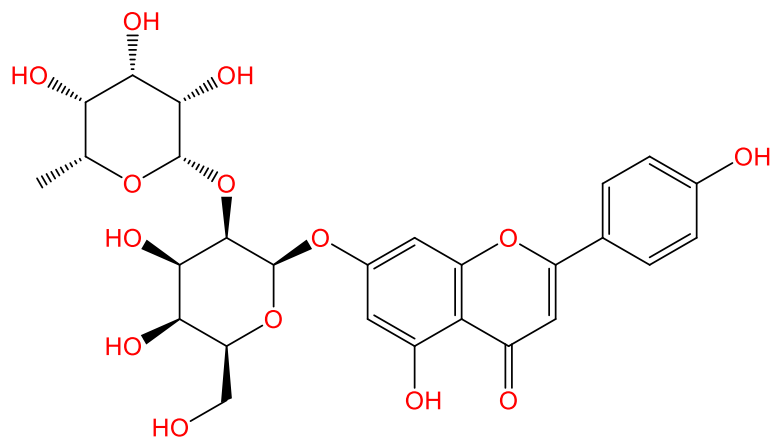 | C[C@H]1O[C@@H](O[C@H]2[C@@H](Oc3cc(O)c4c(=O)cc(-c5ccc(O)cc5)oc4c3)O[C@@H](CO)[C@@H](O)[C@H]2O)[C@@H](O)[C@@H](O)[C@H]1O | -12.238 |
| ZINC4534058 | 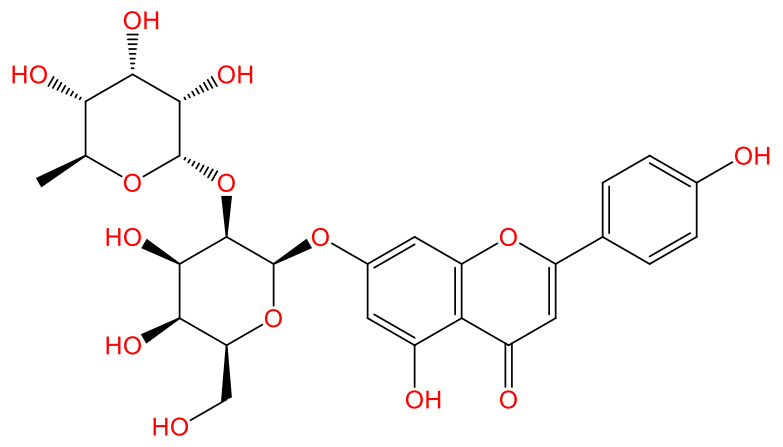 | C[C@@H]1O[C@@H](O[C@H]2[C@@H](Oc3cc(O)c4c(=O)cc(-c5ccc(O)cc5)oc4c3)O[C@@H](CO)[C@@H](O)[C@H]2O)[C@@H](O)[C@@H](O)[C@H]1O | -13.697 |
| ZINC4534059 | 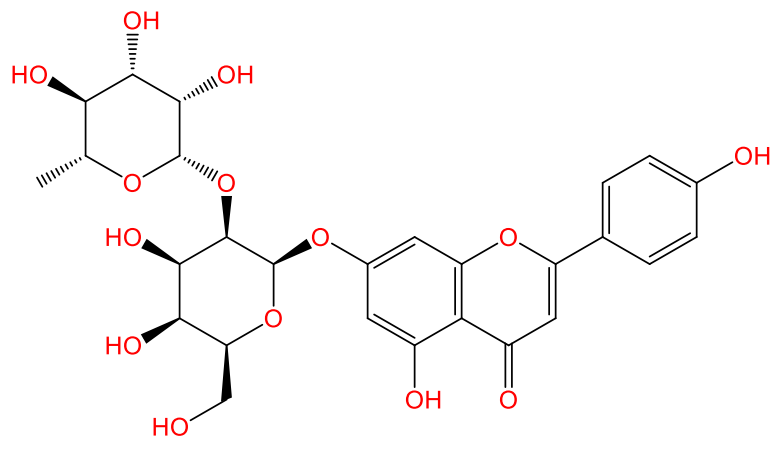 | C[C@H]1O[C@@H](O[C@H]2[C@@H](Oc3cc(O)c4c(=O)cc(-c5ccc(O)cc5)oc4c3)O[C@@H](CO)[C@@H] | -14.387 |
| | | (O)[C@H]2O)[C@@H](O)[C@@H](O)[C@@H]1O | |
| ZINC4534060 | 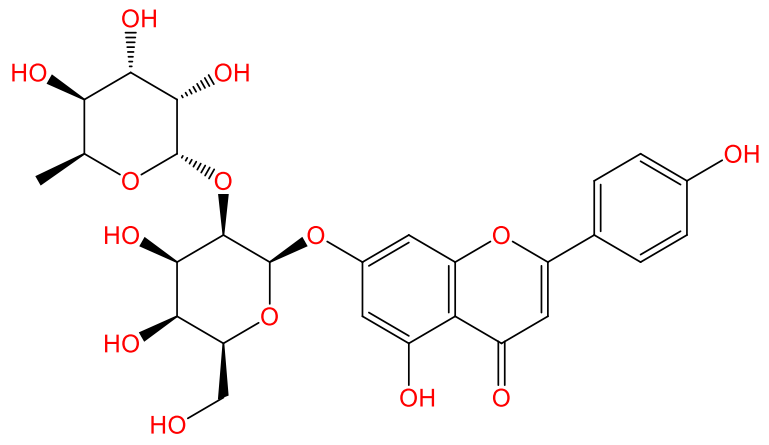 | C[C@@H]1O[C@@H](O[C@H]2[C@@H](Oc3cc(O)c4c(=O)cc(-c5ccc(O)cc5)oc4c3)O[C@@H](CO)[C@@H](O)[C@H]2O)[C@@H](O)[C@@H](O)[C@@H]1O | -11.160 |
| ZINC3978800 (Rhoifolin) | 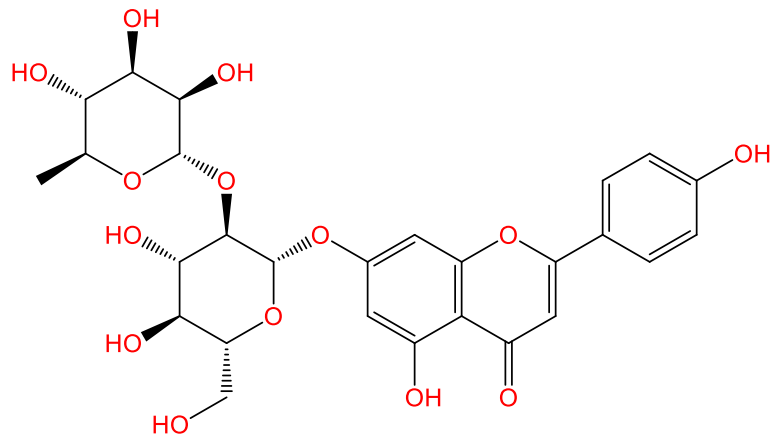 | C[C@@H]1O[C@@H](O[C@H]2[C@H](Oc3cc(O)c4c(=O)cc(-c5ccc(O)cc5)oc4c3)O[C@H](CO)[C@@H](O)[C@@H]2O)[C@H](O)[C@H](O)[C@H]1O | -11.961 |

## Supplemental Table S4

*ADME characteristics of 16 compounds from the ZINC database for natural products, evaluated with the SwissADME resource.*[64]

See SwissADME.xlsx file

# Supplemental Figure S1

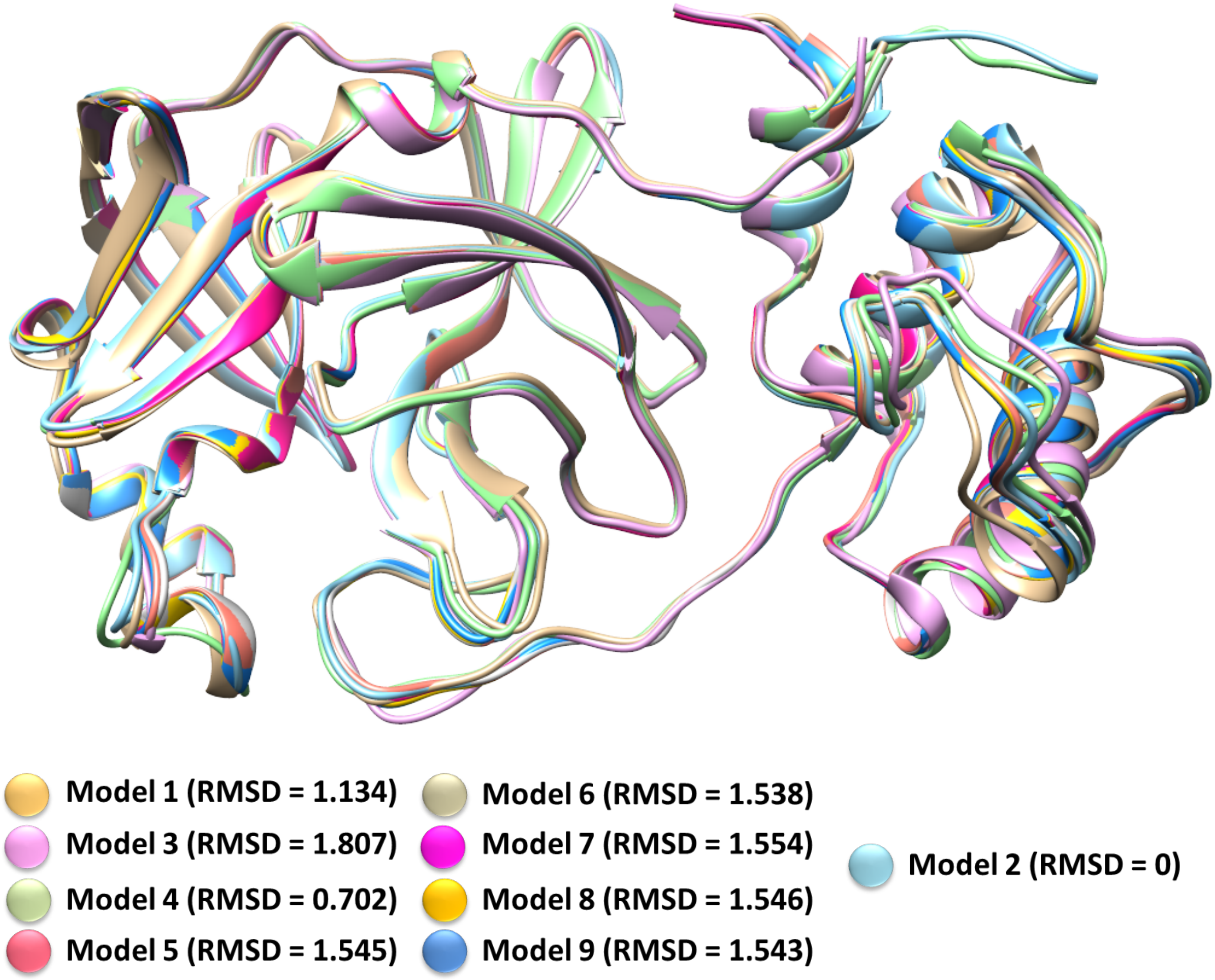

Superposition of 10 different 3CL-PRO models, from the PDB database. RMSDs were calculated with the Chimera program.[14]

## Supplemental Figure S2

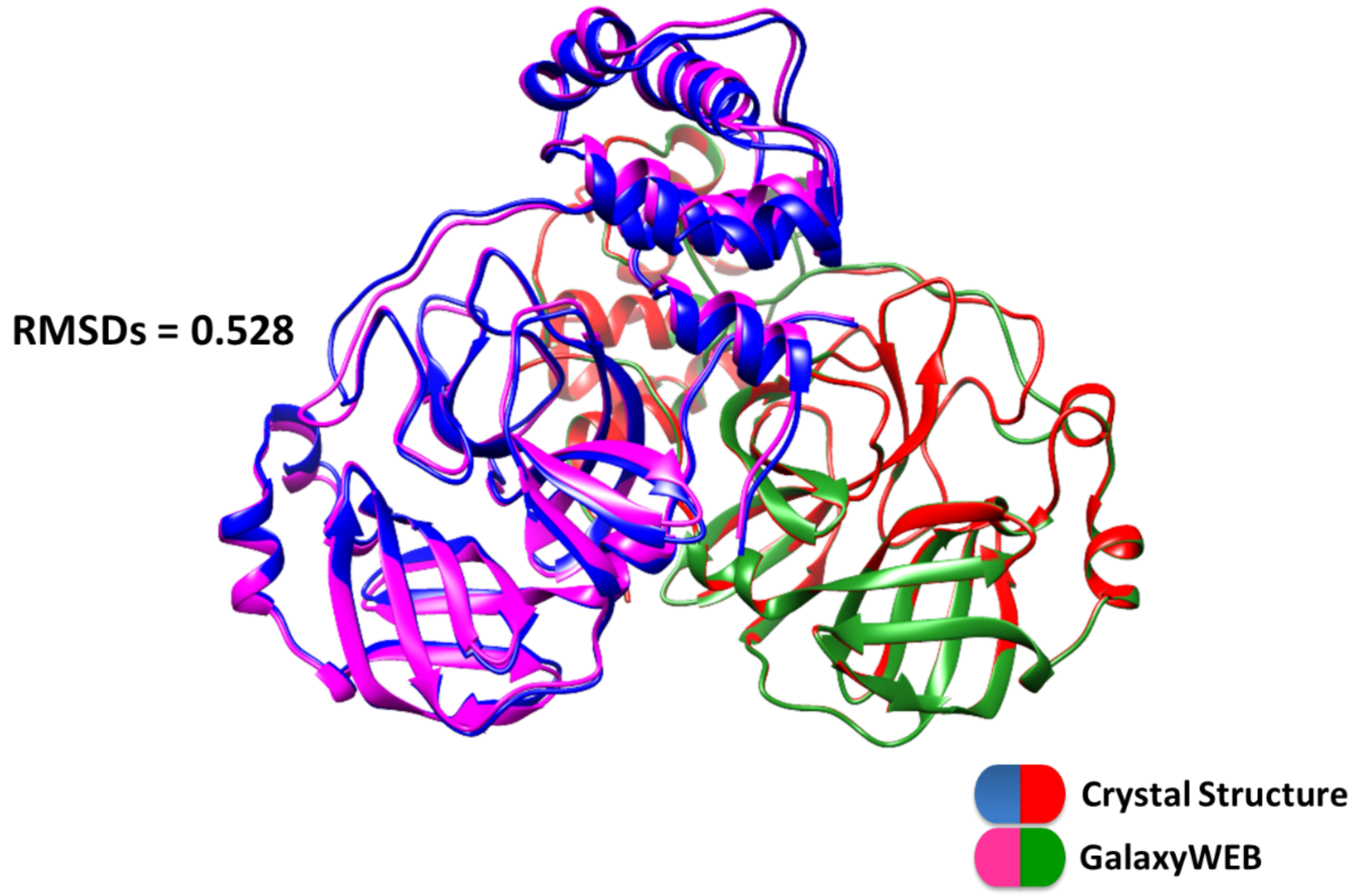

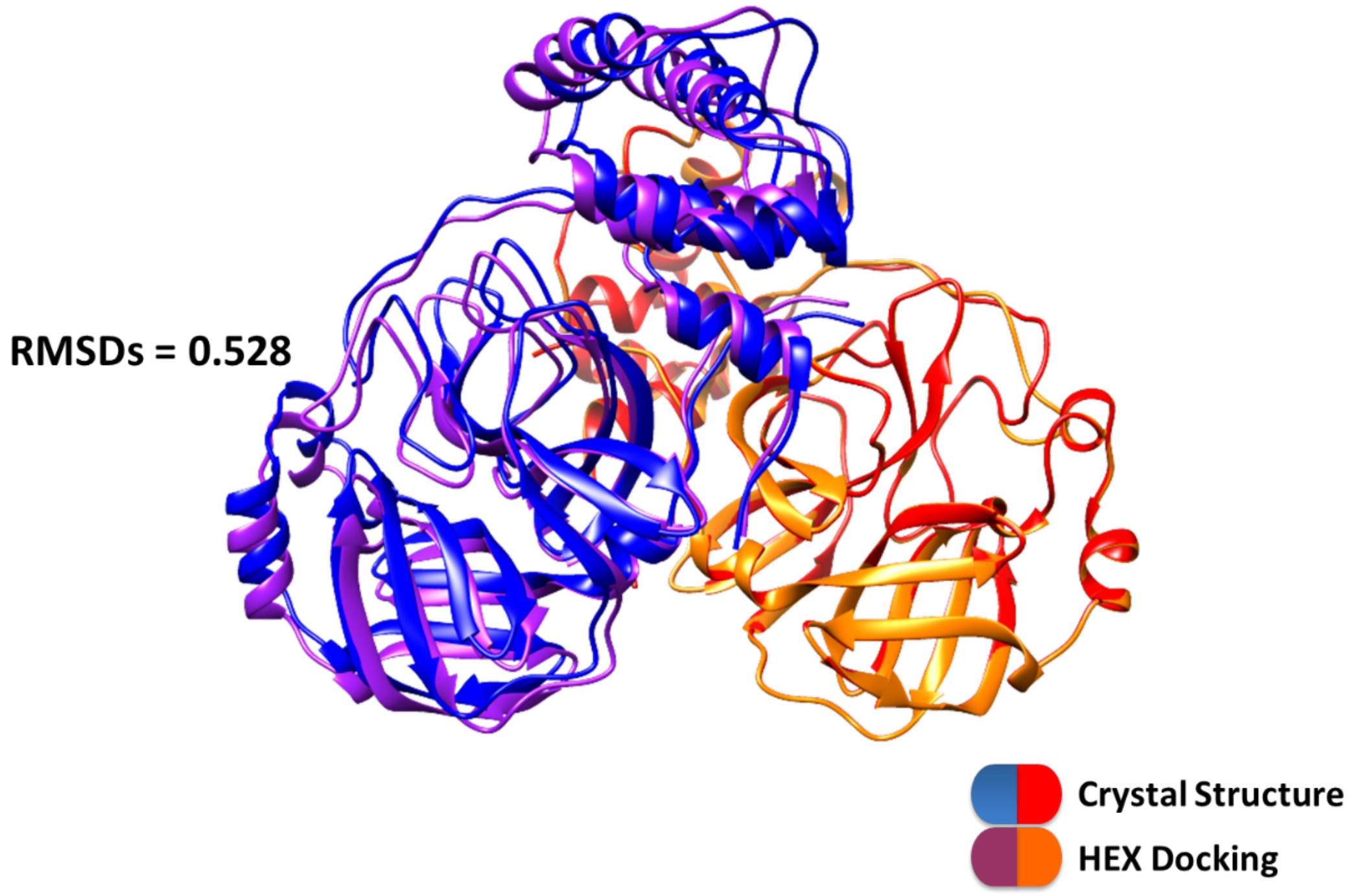

Comparison of the crystal structure (pdb 6YB7) with the two simulated solutions obtained with the GalaxyWeb (routine homomer) [5] and the HEX program.[20,21] The distinct colors in each Figure shows the corresponding monomers.

# SwissADME.xlsx

| Molecule | Smiles | Canonical SMILES | Formula | MW | #Heavy atoms | #Aromatic heavy atoms | Fraction Csp3 | #Rotatable bonds | #H-bond acceptors | #H-bond donors | MR | TPSA |
|---|---|---|---|---|---|---|---|---|---|---|---|---|
| **ZINC4349204 (Fortunellin)** | C[C@@H]1[C@@H]([C@@H]([C@H | OC[C@@H]1O[C@@H](Oc2cc(O)c3c(c2)oc(cc3=O)c2ccc(cc2)OC)[C@@H]([C@@H]([C@@H]1O)O)O | C28H32O14 | 592.55 | 42 | 16 | 0.46 | 7 | 14 | 7 | 141.80 | 217.97 |
| ZINC4349029 | c1cc(ccc1c2cc(=O)c3c(cc(cc3o2)O[C | OC[C@@H]1O[C@@H](Oc2cc(O)c3c(c2)oc(cc3=O)c2ccc(cc2)O)[C@@H]([C@H]([C@@H]1O)O)O[C@ | C26H28O14 | 564.49 | 40 | 16 | 0.42 | 7 | 14 | 8 | 132.56 | 228.97 |
| ZINC4349031 | c1cc(ccc1c2cc(=O)c3c(cc(cc3o2)O[C | OC[C@H]1O[C@@H](Oc2cc(O)c3c(c2)oc(cc3=O)c2ccc(cc2)O)[C@@H]([C@@H]([C@@H]1O)O)O[C@ | C26H28O14 | 564.49 | 40 | 16 | 0.42 | 7 | 14 | 8 | 132.56 | 228.97 |
| ZINC4349034 | c1cc(ccc1c2cc(=O)c3c(cc(cc3o2)O[C | OC[C@H]1O[C@@H](Oc2cc(O)c3c(c2)oc(cc3=O)c2ccc(cc2)O)[C@@H]([C@H]([C@@H]1O)O)O[C@@ | C26H28O14 | 564.49 | 40 | 16 | 0.42 | 7 | 14 | 8 | 132.56 | 228.97 |
| ZINC3983878 | c1cc(ccc1c2cc(c-3c(=O)cc(cc3o2)O[C | OC[C@H]1O[C@@H](Oc2cc(=O)c3c(c2)oc(cc3[O-])c2ccc(cc2)O)[C@@H]([C@H]([C@@H]1O)O)O[C@ | C26H27O14 | 563.48 | 40 | 16 | 0.42 | 7 | 14 | 7 | 129.79 | 231.80 |
| ZINC4349207 | C[C@H]1[C@@H]([C@@H]([C@H]( | OC[C@@H]1O[C@@H](Oc2cc(O)c3c(c2)oc(cc3=O)c2ccc(cc2)OC)[C@@H]([C@@H]([C@@H]1O)O)O | C28H32O14 | 592.55 | 42 | 16 | 0.46 | 7 | 14 | 7 | 141.80 | 217.97 |
| ZINC4349211 | C[C@@H]1[C@@H]([C@H]([C@H]( | OC[C@@H]1O[C@@H](Oc2cc(O)c3c(c2)oc(cc3=O)c2ccc(cc2)OC)[C@@H]([C@@H]([C@@H]1O)O)O | C28H32O14 | 592.55 | 42 | 16 | 0.46 | 7 | 14 | 7 | 141.80 | 217.97 |
| ZINC4349214 | C[C@H]1[C@@H]([C@H]([C@H]([C | OC[C@@H]1O[C@@H](Oc2cc(O)c3c(c2)oc(cc3=O)c2ccc(cc2)OC)[C@@H]([C@@H]([C@@H]1O)O)O | C28H32O14 | 592.55 | 42 | 16 | 0.46 | 7 | 14 | 7 | 141.80 | 217.97 |
| ZINC4349623 | C[C@@H]1[C@@H]([C@@H]([C@H | OC[C@@H]1O[C@@H](Oc2cc(O)c3c(c2)oc(cc3=O)c2ccc(cc2)O)[C@@H]([C@@H]([C@@H]1O)O)O[ | C27H30O14 | 578.52 | 41 | 16 | 0.44 | 6 | 14 | 8 | 137.33 | 228.97 |
| ZINC4349627 | C[C@H]1[C@@H]([C@@H]([C@H]( | OC[C@@H]1O[C@@H](Oc2cc(O)c3c(c2)oc(cc3=O)c2ccc(cc2)O)[C@@H]([C@@H]([C@@H]1O)O)O[ | C27H30O14 | 578.52 | 41 | 16 | 0.44 | 6 | 14 | 8 | 137.33 | 228.97 |
| ZINC4349630 | C[C@@H]1[C@@H]([C@H]([C@H]( | OC[C@@H]1O[C@@H](Oc2cc(O)c3c(c2)oc(cc3=O)c2ccc(cc2)O)[C@@H]([C@@H]([C@@H]1O)O)O[ | C27H30O14 | 578.52 | 41 | 16 | 0.44 | 6 | 14 | 8 | 137.33 | 228.97 |
| ZINC4349633 | C[C@H]1[C@@H]([C@H]([C@H]([C | OC[C@@H]1O[C@@H](Oc2cc(O)c3c(c2)oc(cc3=O)c2ccc(cc2)O)[C@@H]([C@@H]([C@@H]1O)O)O[ | C27H30O14 | 578.52 | 41 | 16 | 0.44 | 6 | 14 | 8 | 137.33 | 228.97 |
| ZINC4534057 | C[C@@H]1[C@@H]([C@@H]([C@@ | OC[C@@H]1O[C@H](Oc2cc(O)c3c(c2)oc(cc3=O)c2ccc(cc2)O)[C@@H]([C@@H]([C@@H]1O)O)O[C@ | C27H30O14 | 578.52 | 41 | 16 | 0.44 | 6 | 14 | 8 | 137.33 | 228.97 |
| ZINC4534058 | C[C@H]1[C@@H]([C@@H]([C@@H | OC[C@@H]1O[C@H](Oc2cc(O)c3c(c2)oc(cc3=O)c2ccc(cc2)O)[C@@H]([C@@H]([C@@H]1O)O)O[C@ | C27H30O14 | 578.52 | 41 | 16 | 0.44 | 6 | 14 | 8 | 137.33 | 228.97 |
| ZINC4534059 | C[C@@H]1[C@H]([C@@H]([C@@H | OC[C@@H]1O[C@H](Oc2cc(O)c3c(c2)oc(cc3=O)c2ccc(cc2)O)[C@@H]([C@@H]([C@@H]1O)O)O[C@ | C27H30O14 | 578.52 | 41 | 16 | 0.44 | 6 | 14 | 8 | 137.33 | 228.97 |
| ZINC4534060 | C[C@H]1[C@H]([C@@H]([C@@H]( | OC[C@@H]1O[C@H](Oc2cc(O)c3c(c2)oc(cc3=O)c2ccc(cc2)O)[C@@H]([C@@H]([C@@H]1O)O)O[C@ | C27H30O14 | 578.52 | 41 | 16 | 0.44 | 6 | 14 | 8 | 137.33 | 228.97 |
| ZINC3978800 | C[C@H]1[C@@H]([C@H]([C@H]([C | OC[C@H]1O[C@@H](Oc2cc(O)c3c(c2)oc(cc3=O)c2ccc(cc2)O)[C@@H]([C@H]([C@@H]1O)O)O[C@@ | C27H30O14 | 578.52 | 41 | 16 | 0.44 | 6 | 14 | 8 | 137.33 | 228.97 |

| | |
|---|---|
| **Fraction Csp3** | The ratio of sp3 hybridized carbons over the total carbon count of the molecule |
| **MR** | Molecular refractivity |
| **TPSA** | Topological polar surface area |
| **iLOGP** | Efficient Description of n-Octanol/Water Partition Coefficient |
| **XLOGP3** | XLOGP3 predicts the logP value of a query compound by using the known logP value of a reference compound as a starting point. |
| **WLOGP** | Lipophilicity factor developed by Wildman and Crippen |
| **MLOGP** | Octanol/water partition coefficient developed by Moriguchi and Matsushita |
| **Silicos-IT Log P** | SILICOS-IT is the log Po/w estimation returned by executing the FILTER-IT program |
| **Consensus Log P** | Consensus log Po/w value is the arithmetic mean of the five predictive values above |
| **ESOL Log S** | Solubility, log S (calculated with the ESOL model) - Estimating Aqueous Solubility Directly from Molecular Structure |
| **ESOL Solubility (mg/ml)** | Estimating Aqueous Solubility Directly from Molecular Structure in mg/ml |
| **ESOL Solubility (mol/l)** | Estimating Aqueous Solubility Directly from Molecular Structure in mol/l |
| **ESOL Class** | Classification of the compound based on estimating Aqueous Solubility Directly from Molecular Structure |
| **Ali Log S** | In silico prediction of aqueous solubility incorporating the effect of topographical polar surface area by Ali et all. |
| **Ali Solubility (mg/ml)** | Aqueous Solubility developed by Ali in mg/ml |
| **Ali Solubility (mol/l)** | Aqueous Solubility developed by Ali in mol/l |
| **Ali Class** | Classification of the compound based on estimating Aqueous Solubility by Ali |
| **Silicos-IT LogSw** | Intrinsic water solubility estimated by Wskowwin executing the FILTER-IT program |
| **Silicos-IT Solubility (mg/ml)** | Intrinsic water solubility in mg/ml |
| **Silicos-IT Solubility (mol/l)** | Intrinsic water solubility in mol/l |
| **Silicos-IT class** | Classification of the compound based on the water solubility executing the FILTER-IT program |
| **log Kp (cm/s)** | Value of the skin permeability coefficient |

| iLOGP | XLOGP3 | WLOGP | MLOGP | Silicos-IT Log P | Consensus Log P | ESOL Log S | ESOL Solubility (mg/ml) | ESOL Solubility (mol/l) | ESOL Class | Ali Log S | Ali Solubility (mg/ml) | Ali Solubility (mol/l) | Ali Class | Silicos-IT LogSw | Silicos-IT Solubility (mg/ml) | Silicos-IT Solubility (mol/l) | Silicos-IT class |
|---|---|---|---|---|---|---|---|---|---|---|---|---|---|---|---|---|---|
| 3.12 | 1.05 | -0.80 | -2.76 | -0.62 | 0.00 | -4.00 | 5.99e-02 | 1.01e-04 | Soluble | -5.22 | 3.59e-03 | 6.05e-06 | Moderately soluble | -2.16 | 4.06e+00 | 6.85e-03 | Soluble |
| 1.85 | -0.36 | -1.49 | -3.16 | -0.72 | -0.77 | -2.95 | 6.38e-01 | 1.13e-03 | Soluble | -3.99 | 5.83e-02 | 1.03e-04 | Soluble | -1.92 | 6.83e+00 | 1.21e-02 | Soluble |
| 2.11 | -0.36 | -1.49 | -3.16 | -0.72 | -0.72 | -2.95 | 6.38e-01 | 1.13e-03 | Soluble | -3.99 | 5.83e-02 | 1.03e-04 | Soluble | -1.92 | 6.83e+00 | 1.21e-02 | Soluble |
| 2.16 | -0.36 | -1.49 | -3.16 | -0.72 | -0.71 | -2.95 | 6.38e-01 | 1.13e-03 | Soluble | -3.99 | 5.83e-02 | 1.03e-04 | Soluble | -1.92 | 6.83e+00 | 1.21e-02 | Soluble |
| 2.92 | -1.04 | -1.10 | -3.39 | -0.72 | -0.67 | -2.51 | 1.73e+00 | 3.07e-03 | Soluble | -3.34 | 2.58e-01 | 4.57e-04 | Soluble | -1.92 | 6.82e+00 | 1.21e-02 | Soluble |
| 3.09 | 1.05 | -0.80 | -2.76 | -0.62 | -0.01 | -4.00 | 5.99e-02 | 1.01e-04 | Soluble | -5.22 | 3.59e-03 | 6.05e-06 | Moderately soluble | -2.16 | 4.06e+00 | 6.85e-03 | Soluble |
| 2.94 | 1.05 | -0.80 | -2.76 | -0.62 | -0.04 | -4.00 | 5.99e-02 | 1.01e-04 | Soluble | -5.22 | 3.59e-03 | 6.05e-06 | Moderately soluble | -2.16 | 4.06e+00 | 6.85e-03 | Soluble |
| 2.74 | 1.05 | -0.80 | -2.76 | -0.62 | -0.08 | -4.00 | 5.99e-02 | 1.01e-04 | Soluble | -5.22 | 3.59e-03 | 6.05e-06 | Moderately soluble | -2.16 | 4.06e+00 | 6.85e-03 | Soluble |
| 2.45 | -0.16 | -1.10 | -2.96 | -1.17 | -0.59 | -3.22 | 3.50e-01 | 6.04e-04 | Soluble | -4.19 | 3.70e-02 | 6.40e-05 | Moderately soluble | -1.48 | 1.92e+01 | 3.31e-02 | Soluble |
| 2.03 | -0.16 | -1.10 | -2.96 | -1.17 | -0.67 | -3.22 | 3.50e-01 | 6.04e-04 | Soluble | -4.19 | 3.70e-02 | 6.40e-05 | Moderately soluble | -1.48 | 1.92e+01 | 3.31e-02 | Soluble |
| 2.61 | -0.16 | -1.10 | -2.96 | -1.17 | -0.56 | -3.22 | 3.50e-01 | 6.04e-04 | Soluble | -4.19 | 3.70e-02 | 6.40e-05 | Moderately soluble | -1.48 | 1.92e+01 | 3.31e-02 | Soluble |
| 2.60 | -0.16 | -1.10 | -2.96 | -1.17 | -0.56 | -3.22 | 3.50e-01 | 6.04e-04 | Soluble | -4.19 | 3.70e-02 | 6.40e-05 | Moderately soluble | -1.48 | 1.92e+01 | 3.31e-02 | Soluble |
| 2.45 | -0.16 | -1.10 | -2.96 | -1.17 | -0.59 | -3.22 | 3.50e-01 | 6.04e-04 | Soluble | -4.19 | 3.70e-02 | 6.40e-05 | Moderately soluble | -1.48 | 1.92e+01 | 3.31e-02 | Soluble |
| 2.63 | -0.16 | -1.10 | -2.96 | -1.17 | -0.55 | -3.22 | 3.50e-01 | 6.04e-04 | Soluble | -4.19 | 3.70e-02 | 6.40e-05 | Moderately soluble | -1.48 | 1.92e+01 | 3.31e-02 | Soluble |
| 2.57 | -0.16 | -1.10 | -2.96 | -1.17 | -0.56 | -3.22 | 3.50e-01 | 6.04e-04 | Soluble | -4.19 | 3.70e-02 | 6.40e-05 | Moderately soluble | -1.48 | 1.92e+01 | 3.31e-02 | Soluble |
| 2.20 | -0.16 | -1.10 | -2.96 | -1.17 | -0.64 | -3.22 | 3.50e-01 | 6.04e-04 | Soluble | -4.19 | 3.70e-02 | 6.40e-05 | Moderately soluble | -1.48 | 1.92e+01 | 3.31e-02 | Soluble |
| 1.31 | -0.16 | -1.10 | -2.96 | -1.17 | -0.81 | -3.22 | 3.50e-01 | 6.04e-04 | Soluble | -4.19 | 3.70e-02 | 6.40e-05 | Moderately soluble | -1.48 | 1.92e+01 | 3.31e-02 | Soluble |

| GI absorption | BBB permeant | Pgp substrate | CYP1A2 inhibitor | CYP2C19 inhibitor | CYP2C9 inhibitor | CYP2D6 inhibitor | CYP3A4 inhibitor | log Kp (cm/s) | Lipinski #violations | Ghose #violations | Veber #violations | Egan #violations | Muegge #violations | Bioavailability Score | PAINS #alerts | Brenk #alerts |
|---|---|---|---|---|---|---|---|---|---|---|---|---|---|---|---|---|
| Low | No | Yes | No | No | No | No | No | -9.17 | 3 | 4 | 1 | 1 | 3 | 0.17 | 0 | 0 |
| Low | No | Yes | No | No | No | No | No | -10.00 | 3 | 3 | 1 | 1 | 3 | 0.17 | 0 | 0 |
| Low | No | Yes | No | No | No | No | No | -10.00 | 3 | 3 | 1 | 1 | 3 | 0.17 | 0 | 0 |
| Low | No | Yes | No | No | No | No | No | -10.00 | 3 | 3 | 1 | 1 | 3 | 0.17 | 0 | 0 |
| Low | No | Yes | No | No | No | No | No | -10.48 | 3 | 2 | 1 | 1 | 3 | 0.11 | 0 | 0 |
| Low | No | Yes | No | No | No | No | No | -9.17 | 3 | 4 | 1 | 1 | 3 | 0.17 | 0 | 0 |
| Low | No | Yes | No | No | No | No | No | -9.17 | 3 | 4 | 1 | 1 | 3 | 0.17 | 0 | 0 |
| Low | No | Yes | No | No | No | No | No | -9.17 | 3 | 4 | 1 | 1 | 3 | 0.17 | 0 | 0 |
| Low | No | Yes | No | No | No | No | No | -9.94 | 3 | 4 | 1 | 1 | 3 | 0.17 | 0 | 0 |
| Low | No | Yes | No | No | No | No | No | -9.94 | 3 | 4 | 1 | 1 | 3 | 0.17 | 0 | 0 |
| Low | No | Yes | No | No | No | No | No | -9.94 | 3 | 4 | 1 | 1 | 3 | 0.17 | 0 | 0 |
| Low | No | Yes | No | No | No | No | No | -9.94 | 3 | 4 | 1 | 1 | 3 | 0.17 | 0 | 0 |
| Low | No | Yes | No | No | No | No | No | -9.94 | 3 | 4 | 1 | 1 | 3 | 0.17 | 0 | 0 |
| Low | No | Yes | No | No | No | No | No | -9.94 | 3 | 4 | 1 | 1 | 3 | 0.17 | 0 | 0 |
| Low | No | Yes | No | No | No | No | No | -9.94 | 3 | 4 | 1 | 1 | 3 | 0.17 | 0 | 0 |
| Low | No | Yes | No | No | No | No | No | -9.94 | 3 | 4 | 1 | 1 | 3 | 0.17 | 0 | 0 |
| Low | No | Yes | No | No | No | No | No | -9.94 | 3 | 4 | 1 | 1 | 3 | 0.17 | 0 | 0 |

| Leadlikeness #violations | Synthetic Accessibility |
| --- | --- |
| 1 | 6.45 |
| 1 | 6.08 |
| 1 | 6.08 |
| 1 | 6.08 |
| 1 | 6.28 |
| 1 | 6.45 |
| 1 | 6.45 |
| 1 | 6.45 |
| 1 | 6.33 |
| 1 | 6.33 |
| 1 | 6.33 |
| 1 | 6.33 |
| 1 | 6.33 |
| 1 | 6.33 |
| 1 | 6.33 |
| 1 | 6.33 |
| 1 | 6.33 |